\documentclass[%
reprint,
amsmath,amssymb,
aps,
prapplied,
]{revtex4-2}

\usepackage{graphicx}
\usepackage{dcolumn}
\usepackage{bm}
\usepackage{xcolor}

\newcommand{\ket}[1]{|#1\rangle}

\begin{document}
	
	\preprint{APS/123-QED}
	
	\title{Simulating spin chains using a superconducting circuit: \\ gauge invariance, superadiabatic transport, and broken time-reversal symmetry}
	
	\author{A. Veps\"al\"ainen}
	
	
	\author{G.~S. Paraoanu}
	\affiliation{QTF Centre of Excellence, Department of Applied Physics, Aalto University School of Science, P.O. Box 15100, FI-00076 AALTO, Finland}
	
	
	\date{\today}
	
	\begin{abstract}
Simulation of materials by using quantum processors is envisioned to be a major direction of development in quantum information science. Here we exploit the mathematical analogies between a triangular spin lattice with Dzyaloshinskii-Moriya coupling on one edge and a three-level system driven by three fields in a loop configuation to emulate spin-transport effects. We show that the spin transport efficiency, seen in the three-level system as population transfer, is enhanced when the conditions for superadiabaticity are satisfied. We demonstrate experimentally that phenomena characteristic to spin lattices due to gauge invariance, non-reciprocity, and broken time-reversal symmetry can be reproduced in the three-level system.

	\end{abstract}
	
	\maketitle
	
	
	\section{Introduction}

Richard Feynman, in a landmark paper from 1982 \cite{Feynman82}, suggested that quantum phenomena might be efficiently predicted by using other, better controllable quantum systems, as simulators.  Later in 1999 Seth Lloyd showed \cite{Lloyd96} that a universal quantum computer based on the gate model \cite{NielsenChuang} can be used to solve the Schr\"odinger equation by the trotterization of its unitary evolution operator. With superconducting qubits, demonstrations of such digital simulations of spin systems have been recently realized \cite{Salathe2015,Barends2016}. While large-scale quantum computers based on the discrete gate model are still decades away, analog simulations on small-scale quantum ``emulators'' are  presently feasible \cite{Paraoanu2014}. The overhead, in terms of number of qubits and operations, is remakably small. For example, single-device multilevel systems such as the one used in the present work have been already utilized for the simulation of large spins \cite{Neeley09}, two-qubit dynamics \cite{Svelitsky14}, and motional averaging \cite{Li2013}.

Here we use a three-level system to simulate  transport phenomena in three-spin chains with XX nearest-neighbour interaction and Dzyalozhinskii-Moriya next-nearest-neighbour interaction. These types of spin lattices play an essential role in our understanding of magnetic phenomena: they contain both the standard XX couplings and the asymmetric spin exchange found phenomenologically by Dzyaloshinskii \cite{Dzialozinskii1957}, and whose microscopic basis, related to spin-orbit coupling and inversion symmetry breaking, has 
been revealed by Moryia \cite{Moryia1960}.
These models have been studied intensively in connection with magnetic phenomena \cite{Huang2017,Kargarian2009,Ma2011,Jafari2018}, see review by \cite{Song2014}. Engineered systems that realize the same spin physics have been proposed in circuit QED \cite{Koch2010} and later realized experimentally \cite{Roushan2017,Scully2019}. Related devices displaying non-reciprocality and broken time-reversal symmetry have been realized in nanomechanics \cite{Laure2018,Peterson2017,Kippenberg2017,Fink2017} and in degenerate ultracold gases \cite{Aidelsburger2013}.

We show that, in general, the spin Hamiltonian maps onto that of a multilevel system with driven transitions with complex matrix elements; thus, a multilevel system can be seen as a universal simulator of spin chains with any type of interaction. We put in evidence effects  such as gauge invariance, chirality, broken-time reversal symmetry, and non-reciprocity. Our focus is on simulating transport phenomena in spin chains by a specific modulation of the couplings which will be discussed in detail below. 
We emphasize that also the imperfections of the real condensed-matter system (inhomogenous broadening in our case) are directly emulated by the multilevel system (through the presence of ac Stark shifts), see also the discussion in Lloyd's seminal paper \cite{Lloyd96}. Thus,  in contrast to the case of digital simulation or quantum information processing, we do not aim at realizing high-fidelity transfer protocols; instead, we are interested in protocols that are demonstrably robust under experimental errors with realistic devices.


In general, there are two ways in which transport of excitations can be realized: sequential and adiabatic. The first implies transferring the excitation between next-neighbour sites by using Rabi pulses \cite{Rabi37}. The sequential method is fast but at the same time sensitive  to errors in the timing of the pulses and their shape.  In contrast, the adiabatic method is based on the modulation of the coupling elements in such a way that the system follows the dark state, and yields the desired robustness against imperfections of the pulses. However, the method is also slow, as required by the adiabatic theorem \cite{Born28}. Several acronyms are used to describe various versions of this process \cite{Menchon_Enrich_2016}. SAP (stimulated adiabatic passage) is a general term encompassing many physical realizations: for example Bose-Einstein condensates in three wells formed by optical trapping, quantum dots, sound waves, coupled waveguides {\it etc.}. Similarly, coherent tunneling via adiabatic passage (CTAP) is often used in works on spin-1/2 particles \cite{Oh2013}, electrons in triple quantum dots \cite{Greentree2004}
and three-well Bose-Hubbard systems \cite{Bradly2012}, and triangular harmonic-trap lattices where single atoms are transferred \cite{Menchon-Enrich2014}. Exactly solvable models of coherent transfer by adiabatic passage in two-dimensional lattices, including triangular ones, were studied in \cite{Longhi2014}. In the specific case of spin lattices however, spatial transport of spins is often refered to as DSAP (dark-state adiabatic passage) \cite{Bergmann2019}, which is the terminology we will also use.

While both the sequential and adiabatic methods have advantages and disadvantages, there exists, surprisingly, a way to get the best of the two worlds. This is based on a simple but powerful observation made by Berry \cite{Berry09} and anticipated by several authors \cite{PingAo2000,Demirplak03,Demirplak05,Demirplak08}: a system can follow exactly the adiabatic state by using an additional counterdiabatic Hamiltonian tailored to cancel the nonadiabatic excitations. This type of evolution is called superadiabatic or transitionless, and several variations have been explored theoretically \cite{Torrontegui13}. In spin systems, transport assisted by counteradiabatic terms has been proposed in \cite{Benseny2017}, a method that can be called superadiabatic DSAP (saDSAP).

In the present experiment, the goal is to simulate this form of transport by using the first three states of a superconducting transmon circuit \cite{transmon_PRA2007,cqed_strong_coupling_wallraff} by controling the system with three microwave tones. This type of driving, called loop-drive or $\Delta$ configuration, has been discussed theoretically in various contexts in atomic physics \cite{Carroll:88,Fleischhauer99,Unanyan97,Pope2019}. Two of  the drives realize the stimulated Raman adiabatic passage (STIRAP), while the third provides the counterdiabatic correction  Hamiltonian required in saDSAP. This configuration results in the creation of a synthetic gauge potential with a gauge-invariant Aharonov-Bohm phase, which can be controlled externally, allowing us to simulate the related gauge-invariance phenomenon in spin systems. This contrasts to the simpler case of two-field drive, where the phases of the driving fields can be eliminated by a gauge transformation, and also with the case of two-level systems, where again the phase of the counterdiabatic pulse is irrelevant. In three-level systems one can use this pulse sequence to realize the superadiabatic STIRAP (saSTIRAP), provided that active time-domain compensation for ac Stark sfifts is performed \cite{DiStefano2016,Vepsalainen2018,Vepsalainen2019}. The results present here show that it is possible to have significant population transfer also in the absence of this technique, simulating the transport in spin chains in the realistic experimental conditions when the presence of energy shifts due to magnetic fields or shifts due to modulation of the couplings.


Our results open up several interesting perspectives in circuit quantum electrodynamics, for example toward the use of driven three-level systems for realizing qubits immune to noise \cite{Retzker2016}. The two-photon driving technique might be useful also 
in other systems which have a forbidden direct transition, for example in quantum optics where the Laporte rule prevents the coupling of levels with the same parity in centrosymmetric molecules. Scaling up to chains of transmons would allow the use of the energy levels as additional synthetic dimensions and the creation of 
synthetic gauge potentials \cite{Goldman:2014,Dalibard:2011}. In such configurations one could perform simulations of field theories governed by the SU(3) gauge symmetry \cite{Zoller2013}, such as lattice QCD with its associated SU(3) color gauge. The special counterdiabatic coupling allows also for various spin-1 particle adiabatic dynamics, realizing the multilevel Cook-Shore model for spins \cite{Cook1979,Hioe:87}. Finally, the three-level transmon can be operated with well-defined detunings, which allows the simulation of detrapping phenomena in small quantum networks \cite{Pope2019}.

In general, superadiabatic methods form a bridge between the two paradigms of quantum control, and allow one to exploit the advantages of both. The combination of robustness under parameter fluctuations and drive errors, together with fast operation times would make superadiabatic protocols especially advantageous for reducing the effects of decoherence and increasing the signal-to-noise ratio. For adiabatic quantum computers \cite{Farhi01}, quantum-annealing processors \cite{annealing,dwave}, and holonomic quantum computing \cite{fastholonomic,holonomic,nonabelianwallraff} this would be one important route to achieving quantum advantage \cite{troyer}. In quantum thermodynamics, during the adiabatic cycle of a quantum engine the system should not only be decoupled from the thermal reservoir but also interlevel transitions should be suppressed, leading to superadiabatic engines with increased power \cite{Chotorlishvili16}, and providing novel insights into the foundations of the third law of thermodynamics \cite{B816102J,0295-5075-85-3-30008,PhysRevA.82.053403}. In cyclic processes such as those used in heat engines, superadiabaticity provide a quantitative expression of Carnot's formulation of the third law of thermodynamics by showing why absolute zero is not achievable in finite time \cite{B816102J,0295-5075-85-3-30008,PhysRevA.82.053403}. Finally, techniques of Floquet-engineering of the counterdiabatic term in Ising models \cite{Claeys2019} and of adiabatic transfer of entanglement in quantum dot arrays \cite{Platero2019} and spin lattices with anti-ferromagnetic (Heisenberg) couplings \cite{Groenland2019}, open new avenues for quantum-information tasks in complex lattices.

The paper is organized as follows: we start in Section II by establishing the mathematical equivalence between the single-excitation three-site spin model and the three-level transmon. We also give a straigthforward derivation of the pulse sequence required for superadiabatic transport. In Section III we present a series of technical details on the experiment. The main results on putting in evidence the gauge-invariant phase, the broken time-reversal symmetry and the currents are presented in Section IV. Our conclusions are presented in Section V.

\section{Mapping of spin models into multilevel systems}

\subsection{Spin models}

\begin{figure}[tbp]
	\centering
	\includegraphics[width=0.97\columnwidth]{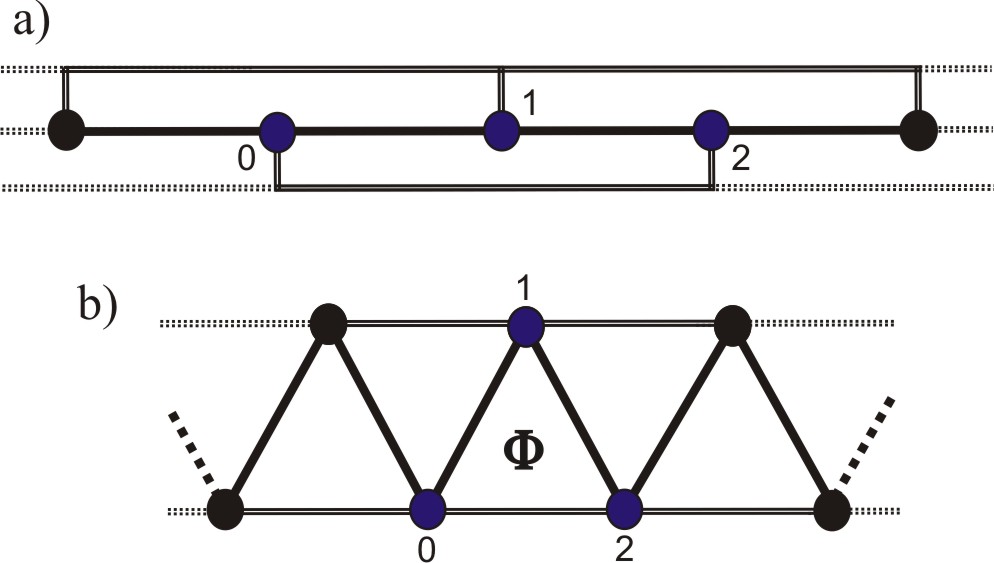}
	\caption{a) A one-dimensional spin lattice with nearest-neighbour (single line) and next-nearest-neighbor (double line) interaction. b) Equivalent two-dimensional representation as a triangular lattice with only nearest-neighbor interaction. In general, the interactions can be complex-valued (Peierls couplings) leading to broken-time reversal symmetry and the development of an Aharonov-Bohm gauge-invariant phase $\Phi$. Three sites, 0,1,2, have been marked here (dark-blue dots), anticipating the mapping to a three-level system.}
	\label{fig:spin_models}
\end{figure}

Our goal is to simulate the transfer of excitation in a spin chain with a structure shown in  Fig. \ref{fig:spin_models}. We consider the spin Hamiltonian in a convenient parametrization, 
\begin{eqnarray}
H &=&\frac{\hbar}{4} \sum_{j \neq k} \Omega_{jk}\cos\phi_{jk}\left(\sigma^{x}_{j}\sigma^{x}_{k} + \sigma^{y}_{j}\sigma^{y}_{k}\right) + \nonumber\\
& &  + \frac{\hbar}{4}\sum_{j \neq k} \Omega_{jk}\sin\phi_{jk}\left(\sigma^{x}_{j}\sigma^{y}_{k} -  \sigma^{y}_{j}\sigma^{x}_{k}\right), \label{eq:spin_hamiltonian}
\end{eqnarray}
which is called the isotropic XX model with Dzyaloshinskii-Moriya interaction. 
The  Dzyaloshinskii-Moriya term is relevant in the proximity of magnetic surfaces where spin-orbit coupling becomes relevant.

We also assume the presence of uncontrolable magnetic fields $B_{j}$ on each site, leading to an additional Zeeman-splitting Hamiltonian which produces inhomogenous broadening 
\begin{equation}
H_{\rm inh} = - \frac{\hbar \gamma}{2} \sum_{j}  B_{j} \sigma_{j}^{z}, \label{eq:inh}
\end{equation}
with $\gamma$ the gyromagnetic ratio. We assume that the $B_{j}$'s are fluctuating around the zero-value.

Here $\sigma^{x}_{j}$ and $\sigma^{y}_{j}$ are the spin-1/2 $x$- and $y$- Pauli matrices associated with the site $j$. Indeed, the first term is the standard XX interaction, symmetric in the exchange of $x$- and $y$ and resulting from the dot product of spins, while the second results from spin-orbit interactions which has the form of a cross-product and it is antisymmetric under the exchange of $x$ and $y$. The parametrization in terms of an angle $\phi$ of the relative strenghts of these interactions has a clear physical meaning if we write $\sigma^{x}_{j} = \sigma^{+}_{j} + \sigma^{-}_{j}$ and $\sigma^{y}_{j} = -i \sigma^{+}_{j} + i \sigma^{-}_{j}$, where 
$\sigma^{\pm}_{j}$ are spin-1/2 raising and lowering operators for site $j$,  $\sigma^{+}_{j}|\uparrow \rangle= |\downarrow \rangle $,  $\sigma^{+}_{j}|\downarrow \rangle= 0 $, 
$\sigma^{-}_{j}|\downarrow \rangle= |\uparrow \rangle$, $\sigma^{-}_{j}|\uparrow \rangle= 0$. Then
\begin{equation}
H = \frac{\hbar}{2}\sum_{j \neq k}\Omega_{jk}\left(e^{i\phi_{jk}}\sigma^{+}_{j}\sigma^{-}_{k} + e^{-i\phi_{jk}}\sigma^{-}_{j}\sigma^{+}_{k}\right). \label{eq:anotherform}
\end{equation}

This type of Hamiltonians are relevant for the analysis of non-trivial spin structures that allow transfer of spin (angular momentum) without transfer of charge \cite{Lima2018,Sentef2007}. The one-dimensional lattice with nearest-neighbour and next-nearest-neighbour interactions from Fig. \ref{fig:spin_models} a) can also be seen as a two-dimensional triangular lattice with only nearest-neighbour interactions, as shown in Fig. \ref{fig:spin_models} b). Such lattices appear in a variety of systems - for example in Bose-Einstein condensates of atoms with two internal states in the Mott-insulator phasem where it yields  three-spin interactions \cite{Pachos2004}. It was shown that spin chains with complex nearest-neighbor and next-nearest neighbor interactions lead to the 
Hofstadter butterfly energy spectrum and to the appearance of edge states
\cite{grass2015}.

Spin transport in this model can be studied  by introducing the spin current operator \cite{Lima2018,Sentef2007}, which is obtained from the continuity equation $\partial_t \sigma_{j}^{z}  + \sum_{k} I_{j\rightarrow k} = 0$.  When comparing it to the Heisenberg equations of motion 
$\partial_t \sigma_{j}^{z}  = \frac{i}{\hbar} [H,\sigma_{j}^{z}]$ we get
\begin{equation}
I_{j\rightarrow k} = i\Omega_{jk}e^{i\phi_{jk}}\sigma^{+}_{j}\sigma^{-}_{k} -i \Omega_{jk}e^{-i\phi_{jk}}\sigma^{-}_{j}\sigma^{+}_{k} .
\end{equation}

Also, the chirality operator in the $z$-direction for the triangular lattice is defined as \cite{Wen1989}
\begin{equation}
C_{z} = \frac{1}{2\sqrt{3}} \vec{\sigma}_{1} \left(\vec{\sigma}_{2} \times  \vec{\sigma}_{3}\right). \label{eq:chirality}
\end{equation}

Consider the three-sites array, which will be the focus of our experiment.
A very useful classification of the eigenstates of the Hamiltonian can be obtained by noticing that $[H,S_{z}] =0$, where $S_{z} = (1/2)\sum_{j}\sigma^{z}_{j}$ is the total spin of the chain in the $z$-direction. Thus, the Hilbert space can be separated in subspaces with $S_{z} = -3/2, -1/2 , 1/2, 3/2$, that is 
$|\downarrow, \downarrow, \downarrow \rangle$, 
$\{$ $|\uparrow, \downarrow, \downarrow \rangle$ , $|\downarrow, \uparrow, \downarrow \rangle $, $|\downarrow, \downarrow, \uparrow \rangle$
$\}$, $\{$ $|\downarrow, \uparrow, \uparrow \rangle$, $|\uparrow, \downarrow, \uparrow \rangle $, $|\uparrow, \uparrow, \downarrow \rangle$
$\}$, and $|\uparrow, \uparrow, \uparrow \rangle$.
In this case, the $S_{z} = -3/2$ and $S_{z}=3/2$ states are left identical by the evolution under the Hamiltonian (\ref{eq:spin_hamiltonian}), while the dynamics on the $S_{z} =-1/2, 1/2$ subspaces correspond to spin waves. These waves can be also seen as the transport of a single excitation (spin-up or spin-down) in the chain.

It is important to realize that the relevant observables do not have cross-couplings between these subspaces. The operator that counts the number of spin excitations, $N = \sum_{j}\sigma^{+}_{j}\sigma^{-}_{j}$
has eigenvalues 0,1,2, and 3 on these subspaces
since $N=3/2 + S_{z}$.
The currents also have zero matrix elements between subspaces with different $S_{z}$.
For the chirality, the eigenvectors $|C_{z}, S_{z}\rangle$ in the subspace $S_{z}=-1/2$ are 
\begin{equation}
|C_{z}, -1/2\rangle = \frac{1}{\sqrt{3}} \left( |\uparrow \downarrow \downarrow \rangle + e^{2iC_{z}\pi /3} |\downarrow \uparrow \downarrow \rangle + 
e^{4iC_{z}\pi /3}|\downarrow \downarrow \uparrow \rangle \right) ,
\end{equation}
with eigenvalues $C_{z} =0, \pm 1$. The eigenvalues of $C_{z}$ on the $S_{z}=1/2$ subspace are obtained by flipping all  the spins, 
\begin{equation}
|C_{z}, 1/2\rangle = \sigma_{1}^{x}\sigma_{2}^{x}\sigma_{3}^{x}|C_{z}, -1/2\rangle .
\end{equation}

\subsection{Multilevel Hamiltonians}

To simulate the dynamics of the Hamiltonian Eq. (\ref{eq:spin_hamiltonian}), the key observation is that the number of spin excitations is conserved by the dynamics. Thus, the $2^3 = 8$ -dimensional Hilbert space breaks down into two 3-dimensional Hilbert spaces and two other additional states with 
no dynamics. Due to this property the simulation can be realized using a three-level system with states $|0\rangle, |1\rangle$ and $|2\rangle$.

Consider for example the subspace $S_{z}=-1/2$ ($N=1$). We can identify $|\uparrow, \downarrow, \downarrow \rangle = (1,0,0)^{T} = |0\rangle$,  $|\downarrow, \uparrow, \downarrow \rangle = (0,1,0)^{T} = |1\rangle$, $|\downarrow, \downarrow, \uparrow \rangle = (0,0,1)^{T} = |2\rangle$. Similarly, for $S_{z}=1/2$ ($N=2$) we identify $|\downarrow, \uparrow, \uparrow \rangle = (1,0,0)^{T} = |0\rangle$,  $|\uparrow, \downarrow, \uparrow \rangle = (0,1,0)^{T} = |1\rangle$, $|\uparrow, \uparrow, \downarrow \rangle = (0,0,1)^{T} = |2\rangle$.

Inhomogenous-broadening terms appear in the simulator mostly as a result of ac Stark shifts, which can be significant if the values of the amplitudes of the pulses are large. We therefore have 
\begin{equation}
H_{\rm ac Stark} = \frac{\hbar}{2} \sum_{j} \epsilon_{j} |j\rangle \langle j| , \label{eq:acStark}
\end{equation}
which reproduces the action of $H_{\rm inh}$ with $\epsilon_{j} = \pm \gamma  \left( \sum_{k\neq j} B_{k} -B_{j} \right)$ on the subspaces with $S_{z} = \pm 1/2$.

The operators appearing in Eq. (\ref{eq:spin_hamiltonian}) can be identified with the Gell-Mann operators (see Supporting Information SI1 A)
\begin{align} 
	\Lambda_{jk}^{s} \leftrightarrow \sigma^{+}_{j}\sigma^{-}_{k} + \sigma^{-}_{j}\sigma^{+}_{k} = \frac{1}{2} \left(\sigma^{x}_{j}\sigma^{x}_{k} +  \sigma^{y}_{j}\sigma^{y}_{k}\right) ,\\
	\Lambda_{jk}^{a} \leftrightarrow -i\sigma^{+}_{j}\sigma^{-}_{k} + i \sigma^{-}_{j}\sigma^{+}_{k} = - \frac{1}{2}\left(\sigma^{x}_{j}\sigma^{y}_{k} -  \sigma^{y}_{j}\sigma^{x}_{k}\right) .\label{eq:Hamiltonian_couplings}
\end{align}
Here $\Lambda_{jk}^{s}$ and $\Lambda_{jk}^{a}$ are the symmetric and antisymmetric Gell-Mann operators defined as:  
$\Lambda_{jk}^{s} =  \Lambda_{kj}^{s} = |j\rangle \langle k| +  |k\rangle \langle j|$ (symmetric) and $\Lambda_{jk}^{a} = - \Lambda_{kj}^{a} =  -i|j\rangle\langle k| + i|k\rangle\langle j|$ (anti-symmetric).

As we will see, the Hamiltonian 
\begin{equation}
H = \frac{\hbar}{2} \Omega_{01} \hat{{\bf n}}_{01}\cdot {\bf \Lambda}_{01} + \frac{\hbar}{2} \Omega_{12} \hat{{\bf n}}_{12}\cdot {\bf \Lambda}_{12}+ \frac{\hbar}{2} \Omega_{02} \hat{{\bf n}}_{02}\cdot {\bf \Lambda}_{02}, \label{eq:full_hamiltonian}
\end{equation}
can be implemented by driving a transmon qubit in the loop configuration with 
Rabi frequencies $\Omega_{jk}= \Omega_{kj}$, and with $\phi_{jk}$  the phases of the driving fields where by convention $\phi_{jk} = - \phi_{kj}$,
where 
$\hat{{\bf n}}_{jk}$ are unit vectors in a plane defined as $\hat{{\bf n}}_{jk} = (\cos\phi_{jk},- \sin\phi_{jk})$, with $j,k \in \{0,1,2 \}$. The matrix vector comprising the Gell-Mann matrices is defined as ${\bf \Lambda}_{jk} = (\Lambda_{jk}^{s}, \Lambda_{jk}^{a})$.

This Hamiltonian realizes the so-called loop driving configuration for three-level systems \cite{Fleischhauer99,Unanyan97} (also referred to as $\Delta$ configuration \cite{Pope2019})
with complex (Peierls) couplings between each pair of states.

In the simulator, the currents can be obtained from identifying the population on a level $j$ with the operator $\frac{1}{2}(1+ \sigma_{z})$ for the case $S_{z}=-1/2$ ($N=1$) and with  $\frac{1}{2}(1-\sigma_{z})$ for the case $S_{z}=1/2$ ($N=2$). Indeed, when averaged on superpositions of $\{$ $|\uparrow, \downarrow, \downarrow \rangle$ , $|\downarrow, \uparrow, \downarrow \rangle $, $|\downarrow, \downarrow, \uparrow \rangle$ $\}$ these operators yield the modulus squared of the complex amplitude of the state with the $j$ spin flipped. Thus, the currents in the simulator are 

\begin{equation}
I_{j \rightarrow k} = -\frac{\Omega_{jk}}{2}\left(\sin \phi_{jk} \Lambda_{jk}^{s} - \cos \phi_{jk} \Lambda_{jk}^{a} \right). \label{currents}
\end{equation}

The chiral operator corresponding to Eq. (\ref{eq:chirality}) can be identified straigthforwardly as 
\begin{equation}
C_{z} = \frac{\sqrt{3}}{3}\left( \Lambda^{a}_{01} + \Lambda^{a}_{12} + \Lambda^{a}_{20} \right). \label{eq:chiralityGellMann}
\end{equation}

Chiral states are obtained as a quantum Fourier transform 
\begin{equation}
|C_{z}\rangle = \frac{1}{\sqrt{3}} \sum_{j = 0,1,2} e^{2\pi ijC_{z}/3}|j\rangle ,
\end{equation}
which can be immediately inverted
\begin{equation}
|j\rangle = \frac{1}{\sqrt{3}} \sum_{j = 0,1,2} e^{-2\pi ijC_{z}/3}|C_{z}\rangle .
\end{equation}

\begin{table}[t!]
	\begin{tabular}{||c||c||} 
		\hline
		Spin chain & Simulator \\ [0.5ex] 
		\hline\hline
		N=1 or N=2 subspaces & qutrit Hilbert space \\ 
		\hline
		XX interaction & $\Lambda^{s}$ coupling \\
		\hline
		Dzyaloshinskii-Moriya interaction & $\Lambda^{a}$ coupling \\
		\hline
		inhomogenous broadening & ac Stark shifts \\	
		\hline
		chirality & quantum Fourier\\
		\hline	
		DSAP & STIRAP \\
		\hline
		saDSAP & saSTIRAP	\\
		\hline			
	\end{tabular}
	\caption{Summary of equivalence between the spin chain and the simulator.}
\end{table}

\subsection{Adiabatic and superadiabatic processes}

The possibility of manipulation the couplings of the spin chain raises the issue of efficient and robust transfer of state between sites. This can be done by employing adiabatic and superadiabatic processes.

For both the spin chain and the multilevel simulator 
we can define the DSAP and respectively the STIRAP processes by the requirement that the system follows the dark state $|\rm D(t)\rangle =  \cos{\Theta (t)}\ket{0} - \sin{\Theta (t)}\ket{2}$, as the mixing angle $\Theta (t) = \tan^{-1} [\Omega_{01}(t)/\Omega_{12}(t)]$ is varied slowly from $0$ to $\pi/2$. Let us recall that the  eigenvalues of the STIRAP Hamiltonian $H_{01}(t)+ H_{12}(t)$ comprise the dark state $|D (t)\rangle$ with eigenvalue $0$, as well as two states of the form $ \left(\sin \Theta |0\rangle + \cos \Theta |2\rangle \pm |1\rangle \right) /\sqrt{2}$ with eigenvalues $\pm \hbar \sqrt{\Omega_{01}^2 + \Omega_{12}^2}/2$ respectively. Here $\sin \Theta |0\rangle + \cos \Theta |2\rangle $ is the bright state, orthogonal to the dark state in the subspace $\{|0\rangle , |2\rangle \}$.

To accelerate the process, one could use the concept of superadiabaticity, where a counterdiabatic correction pulse is applied to suppress excitations on states other than the dark state. The resulting protocols are denoted by saSTIRAP for the simulator and saDSAP for the spin chain. Table I summarizes the  equivalence between the two systems. The form of the counterdiabatic pulse in the case of three-level systems can be found by applying the general superadiabatic protocol \cite{Berry09,Demirplak03,Demirplak05,Demirplak08} to the case of STIRAP \cite{Chen10,Giannelli14}. It  is interesting to note that in this specific case, the counterdiabatic Hamiltonian was found \cite{Fleischhauer99,Unanyan97} several  years before the general formalism \cite{Berry09,Demirplak03,Demirplak05,Demirplak08} was developed.
In Supporting Information SI 1C we provide a proof of these results based on the method of adiabatic potentials \cite{Polkovnikov2017}. Here we give a direct, straightforward derivation based on the Schr\"odinger equation.

Specifically, we would like to find under which conditions the dark state is a solution of the Schr\"odinger equation with total Hamiltonian $H_{01}(t)+ H_{12}(t) + H_{02}(t)$. This leads immediately to 
\begin{equation}
\sin \Theta \left (2i \dot{\Theta} - e^{-i\Phi} \Omega_{02}\right)|0\rangle + \cos \Theta \left (2i \dot{\Theta} + e^{i\Phi} \Omega_{02}\right)|2\rangle = 0 ,
\end{equation}
where $\Phi = \phi_{01} + \phi_{12} + \phi_{20}$ is the gauge-invariant phase, to be discussed in detail later.
We see that this can be satisfied only if $\Phi = - \pi /2$ and $\Omega_{02} = 2 \dot{\Theta}(t)$. Thus, if we set $\phi_{01}= \phi_{12}=0$, the counterdiabatic Hamiltonian takes the form
\begin{equation}
H_{\rm cd}(t) = -\frac{\hbar}{2} \Omega_{02}(t) \Lambda_{02}^{a} = \frac{\hbar}{2} \Omega_{02}(t) \Lambda_{20}^{a}. \label{eq:cd}
\end{equation}
where, as previously, $\Lambda_{kl}^{a} = - \Lambda_{lk}^{a} =  -i|k\rangle\langle l| + i|l\rangle\langle k|$ are the anti-symmetric Gell-Mann matrices.

It is interesting to note that the Gell-Mann matrices 
$\Lambda_{01}^{s}$, $\Lambda_{12}^{s}$, and	$\Lambda_{02}^{a}$, form a closed subalgebra and can be regarded respectively as the $x$, $y$, and $z$-components of a spin-1 particle
since $[\Lambda_{01}^{s}$, $\Lambda_{12}^{s}] = i\Lambda_{02}^{a}$ and circular permutations thereof. Thus, STIRAP can be seen as the adiabatic guiding of a spin-1 in the $xOy$ plane by a magnetic field with $x,y$-components $(\Omega_{01}, \Omega_{12})$. The mixing angle $\Theta$ is then the angle formed by the magnetic field (and the spin which follows its direction) with the $y$-axis.  Interestingly, saSTIRAP achieves a faster motion in the same plane by adding a control field in the $z$-direction: the corresponding spin $z$-component produces a rotation in the $x-y$ plane designed such that it cancels exactly the nonadibatic terms.

\section{Experimental platform} 

\subsection{Measurement and control setup}

\begin{figure*}[tbp]
	\centering
	\includegraphics[width=0.97\textwidth]{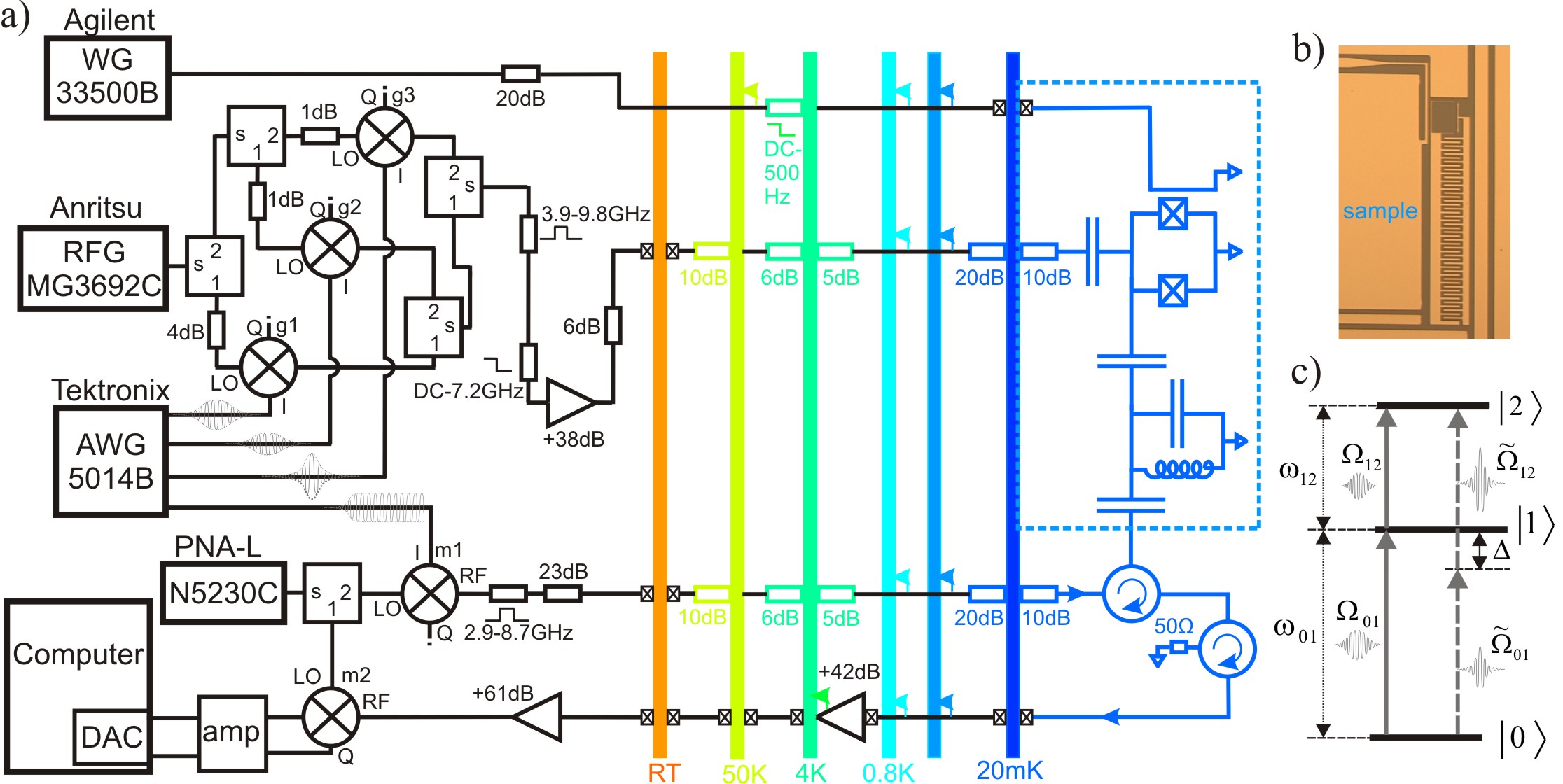}	
	\caption{Schematic of the experimental platform used for simulations. a) Microwave instruments and components at various temperature stages are used to control the transmon device installed at the mixing chamber. b) Optical image of the sample, showing the transmon and part of the coplanar waveguide resonator used for measurement. c) Energy levels and driving fields in the loop driving configuration. STIRAP pulses $\Omega_{01}(t)$ and $\Omega_{02}(t)$ are applied resonantly into the transitions $\omega_{01}(t)$ and $\omega_{02}(t)$ respectively, while the counteradiabatic control is realized by the two pulses $\tilde{\Omega}_{01}$ and $\tilde{\Omega}_{12}$ with detuning $\Delta$.}
	\label{fig:schematic_saSTIRAP}
\end{figure*}

Our experiments run on a superconducting-circuit platform, as shown in Fig. \ref{fig:schematic_saSTIRAP} a). As a multilevel simulator we use  a transmon device \cite{transmon_PRA2007}, which consists  of a Cooper pair box with large shunting capacitors  inserted in the gap between the signal line and the ground of a coplanar waveguide. The latter is configured as a $\lambda /4$ resonator and used for dispersive readout. The bare resonance frequency  of the resonator (measured with the qubit far-detuned) is $f_r \simeq 5.13\ \textrm{GHz}$, and for the quality factor we obtain $Q \simeq 7000$.  The size of the Josephson junctions is $150\times 170\ \textrm{nm}^2$ and it is fabricated from aluminum ($90\ \textrm{nm}$ film thickness) by shadow angle evaporation on a high-resistivity Si substrate; the chip is bonded and installed in a dilution refrigerator with $\sim 20\ \textrm{mK}$ base temperature, see Fig. \ref{fig:schematic_saSTIRAP} b). The device is biased by a magnetic field applied by using an additional line 
which is shortcut to the ground in the proximity of the SQUID loop of the device. For this, at room temperature we use an Agilent 33500B waveform generator, while a passive low pass RC-filter anchored to the 4K-flange of the refrigerator (cut-off frequency of  $\sim 500\ \textrm{Hz}$) is used to filter out the thermal noise. 

The transmon is controlled by sending microwave pulses through a coplanar waveguide which is evaporated on the chip and capacitively interacts with the large transmon shunting capacitor. The pulses are created by mixing their envelopes, created by an arbitrary wavefrom generator (Tektronix 5014B), with a continuos microwave tone. In the setup, three IQ-mixers (IQ-0307L), denoted by g1, g2 and g3, are used to create pulses at three different control frequencies, see Fig. \ref{fig:schematic_saSTIRAP} a). In order to ensure the phase-coherence of the pulses, a single local oscillator tone at 7.608 GHz (generated by an Anritsu MG3692C) is used, and the pulse envelopes are digitally modulated by an intermediate frequency tone.

The mixers are calibrated at the beginning of each experiment by standard techniques in order to reduce the leakage and the spurious sidebands.  The detection scheme is a homodyne measurement: the signal from a vector network analyzer (PNA-L N5230C) at a frequency $f_p = 5.1249\ \textrm{GHz}$ is split into two parts, one mixed in the IQ-mixer m1 (IQ-0307LXP) with a rectangular waveform, and the other used as the LO. Demodulation and digitization is done by mixer m2 (IQ-0307LXP) and by an analog-to-digital converter (Acquiris U1082a). To perform quantum state tomography, we record the demodulated traces in time domain. We first prepare the system in the states $\ket{0}$, $\ket{1}$ and $\ket{2}$ and use these traces as calibration. The calibration measurement fidelities of the states $\ket{1}$ and $\ket{2}$, with $\ket{0}$ as reference, are 95.7\% and 88.4\% respectively. To extract the populations for a general state, we assume the measured trace to be a linear combination of the calibration traces. Using the least squares fit, we can extract the coefficients of the linear combination in the basis of the calibration traces \cite{three_level_tomography}.

We first characterize the device: spectroscopy measurements allow us to identify the transition frequencies
$\omega_{01}$ and $\omega_{12}$ between the energy levels of the device at different bias magnetic fields and extract the parameters of the electrical circuit. We obtain a Josephson energy at the sweet spot $E_{J\Sigma} = E_{J,1} + E_{J,2} \simeq h\times 26.235\ \textrm{GHz}$, a charging energy $E_C \simeq h\times 282\ \textrm{MHz}$ (which results in an anharmonicity $\hbar \omega_{12}-\hbar \omega_{01} \approx - E_{\rm C}$), and a junction asymmetry $d = \lvert E_{J,1} - E_{J,2}\rvert/(E_{J,1} + E_{J,2}) \simeq 0.02$. When the qubit frequency is tuned to be on resonance with the $\lambda/4$ resonator, we observe an avoided crossing in the spectrum, which allows us to extract the qubit-resonator coupling $g \simeq 103\ \textrm{MHz}$. After this, the magnetic field is kept fixed at a bias point  corresponding to $\omega_{01}/(2\pi) = 7.381$ GHz and $\omega_{12}/(2\pi)= 7.099$ GHz. 

At this biasing point, we measure the relaxation rates from the state 1 and 2 by exciting the system with $\pi$ pulses and recording the decay. 
We obtain $\Gamma_{10} =$ 5.0 MHz and $\Gamma_{21} =$ 7.0 MHz. From Ramsey interference experiments, we find that in this sample the dephasing times are dominated by the energy relaxation. To model the decoherence, we use the standard Lindblad formalism for a three-level system \cite{EIT_abdumalikov,ourPRB}, with a superoperator ${\cal L} [\rho ] = -\Gamma_{21} \rho_{22} |2\rangle\langle 2| - (\Gamma_{10} \rho_{11} - \Gamma_{21} \rho_{22}) |1\rangle\langle 1| + \Gamma_{10} \rho_{11} |0\rangle\langle 0|$. 

\vspace{2mm}
{\bf STIRAP pulses}
\vspace{2mm}

In order to create the microwave pulses used for STIRAP, we apply two IF waves with Gaussian envelopes  $\exp\left[-\frac{t^2}{2\sigma^2}\right]$, $\exp\left[-\frac{(t-t_{s})^2}{2\sigma^2}\right]$ and phases $\phi_{01}$, $\phi_{12}$ to the I ports of the mixers g1 and g2. These pulses couple resonantly into the $0 \leftrightarrow 1$ and $1 \leftrightarrow 2$ transitions, resulting in Rabi couplings 
\begin{eqnarray}
\label{eq:gaussian_envelopes}
\Omega_{01}(t) &=& \Omega_{01}\exp\left[-\frac{t^2}{2\sigma^2}\right], \\
\Omega_{12}(t) &=& \Omega_{12}\exp\left[-\frac{(t - t_{s})^2}{2\sigma^2}\right],
\end{eqnarray}
and yielding the matrix elements $\langle 0| H(t) |1\rangle = \langle 1| H(t) |0\rangle^{*} = \Omega_{01}(t) \exp (i \phi_{01})$ and $\langle 1| H(t) |2\rangle = \langle 2| H(t) |1\rangle^{*} = \Omega_{12}(t) \exp (i \phi_{12})$.  For convenience, the Gaussians were truncated at $\pm 3\sigma$.
The Gaussian pulses are not the only possible choice for the STIRAP pulse shape, but they are experimentally and theoretically convenient without sacrificing  performance \cite{Giannelli14}.
In this parametrization $\sigma$ is the width of the pulses, and the counterintuitive STIRAP sequence is realized at negative pulse separation times $t_{\rm s}<0$.
The resulting form of the STIRAP Hamiltonian \cite{stirap_ours} in the Gell-Mann representation is then
\begin{equation}
H_{\rm STIRAP} (t) = \frac{\hbar}{2} \Omega_{01}(t) \hat{{\bf n}}_{01}\cdot {\bf \Lambda}_{01} + \frac{\hbar}{2} \Omega_{12}(t) \hat{{\bf n}}_{12}\cdot {\bf \Lambda}_{12}, \label{eq:hamiltonian_zero} 
\end{equation}
which reproduces the first two terms of Eq. (\ref{eq:full_hamiltonian}). 

\vspace{2mm}
{\bf Counterdiabatic drive}
\vspace{2mm}

If the adiabaticity condition for STIRAP is broken, for example by attempting to drive the system too fast, the system gets diabatically excited away from the state $\ket{D(t)}$, reducing the transferred population. 
However, it is possible to accelerate the STIRAP protocol by employing the technique of reverse Hamiltonian engineering \cite{Berry09,Demirplak03,Demirplak05,Demirplak08}. This requires the addition of a counterdiabatic pulse with a very specific shape and with complex coupling into the $0-2$ transition - which for the case of transmon is a forbidden transition in the first order. To create this pulse, we generate an IF signal with envelope $\cosh^{-1/2}\left[t_{s}(t-t_{s}/2)/\sigma^{2}\right]$ and phase  $\tilde{\varphi}$, and apply it to the I port of the mixer g3.  The frequency of the IF tone is set such that, after mixing with the local oscillator, the resulting upconverted frequency would match half of the forbidden transition $\omega_{02}/(2\pi) = 14.480$ GHz. Thus, this two-photon drive is detuned from both the $0-1$ and $1-2$ transitions by an amount $\pm \Delta$, 
which equals to half the transmon anharmonicity $\Delta = (\omega_{01}-\omega_{12})/(4\pi)= 141$ MHz.
This generates an effective matrix element $\langle 0| H(t) |2\rangle = \langle 2| H(t) |0\rangle^{*} = \Omega_{02}(t) \exp (i \phi_{02})$ without populating the state $|1\rangle$. The Rabi coupling $\Omega_{02}(t)$ is obtained from the perturbation theory\cite{James2007} as $\Omega_{02}(t) = \tilde{\Omega}_{01}(t) \tilde{\Omega}_{12}(t) /(2 \Delta )$ and $\phi_{02} = - \phi_{20} = 2\tilde{\varphi} + \pi$. Note that we define $\Omega_{02}$ as a real positive quantity. Thus we obtain the last term from Eq. (\ref{eq:full_hamiltonian})
\begin{equation}
H_{02}(t) = \frac{\hbar}{2} \Omega_{02}(t) \hat{{\bf n}}_{02}\cdot {\bf \Lambda}_{02}.
\end{equation} 
Satisfying the  relation $\Phi = \phi_{01} + \phi_{12} + \phi_{20} = - \pi/2$ amounts to producing a complex Peierls matrix element 
$\langle 0 | H_{\rm cd} (t)| 2 \rangle = \langle 2 | H_{\rm cd} (t)| 0 \rangle^{*} = (\hbar /2) \Omega_{02}(t) \exp (i \pi/2)$.  For equal-amplitude Gaussian STIRAP pulses, from the relation $\Omega_{02}(t) = 2 \dot{\Theta} (t)$ we get $\Omega_{02} (t) = - (t_{s}/\sigma^2)\cosh^{-1}\left[t_{s}(t-t_{s}/2)/\sigma^{2}\right]$.

\subsection{Pulse calibration}

Overall, the pulses described above results in couplings of the form $H(t) = \sum_{j \neq k} H_{jk}(t)$, where $H_{jk}(t) = \frac{\hbar}{2} \Omega_{jk}(t) \left( \cos\phi_{jk} \Lambda_{jk}^{s} - \sin\phi_{jk} \Lambda_{jk}^{a} \right)$, which reproduces the form Eq. (\ref{eq:full_hamiltonian}). In addition to these terms, ac Stark shifts are produced by off-resonant drives. In our case, the largest ac Stark shifts are produced by the two-photon pulse, which effectively displaces the energy levels of the qutrit as seen by the STIRAP pulses; this produces in general (see Supporting Information SI2) 
$H_{\rm ac Stark}(t) = \frac{\hbar}{2} \sum_{j} \epsilon_{j}(t) |j\rangle \langle j|$ as in Eq. (\ref{eq:acStark}). These shifts are expected to occur also in the spin chain, where they will appear as inhomogenous broadening. In principle it is possible to exactly cancel these shifts by techniques such as time-dependent frequency corrections \cite{DiStefano2016,Vepsalainen2018,Vepsalainen2019} or by an additional two-photon drive, designed with a detuning with opposite sign and a $\pi$ phase in one of the drives \cite{Retzker2016}. However, whether these techniques can be implemented depends on the particular physical system and the associated array of available experimental methods. For example, in optical systems the control of the phase of the lasers might not be possible with sufficient accuracy. 

\begin{figure}[ht]
	\includegraphics[width=0.45\textwidth]{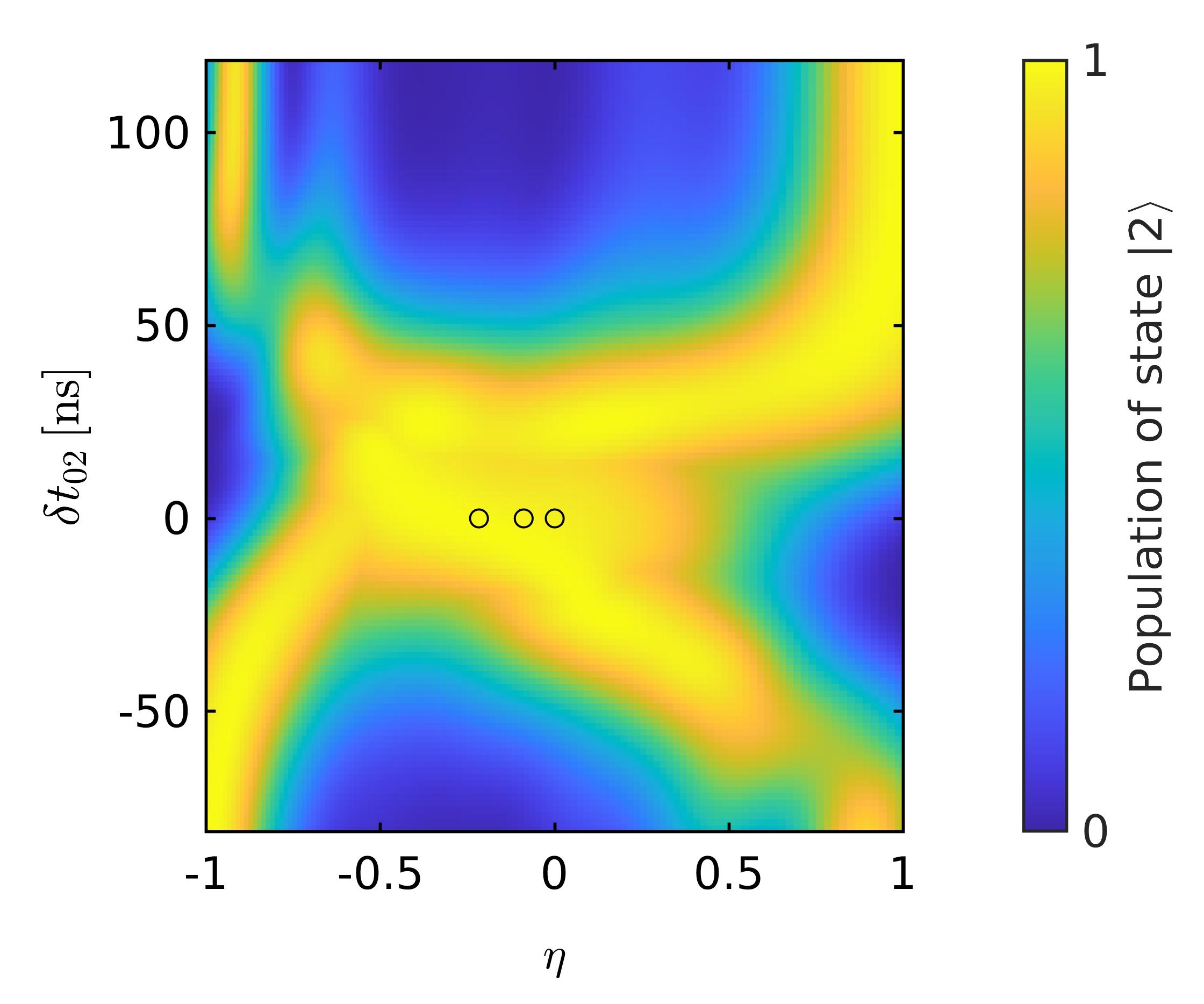}	
	\caption{The population of the state $\ket{2}$ as a function of the amplitude pulse asymmetry $\eta = (\Omega_{12} - \Omega_{01})/(\Omega_{12} +\Omega_{01})$ of the STIRAP pulses and 
		of the time shifts of the counterdiabatic pulse $\Omega_{\rm 02} (t) \rightarrow \Omega_{\rm 02} (t-\delta t_{02})$. The black circles correspond to the values used in experiments, $\eta = -0.22,-0.09$ and a reference $\eta =0$, from left to right. The parameters used in the numerics are $k= |t_{\rm s}|/\sigma = 2.45$ and $\sigma = 25$ ns. 
	}
	\label{fig:calibrations}
\end{figure}

Here we attempt to optimize the transfer by varying the parameters of the STIRAP pulses and the timing of the counterdiabatic pulse. The results are presented in Fig. \ref{fig:calibrations}.  Note that for the numerics we use the full Hamiltonian $\mathcal{H} = \mathcal{H}_{01} + \mathcal{H}_{12} + \mathcal{H}_{02}$ as given in Eq. (\ref{H01}, \ref{H12}) and Eq. (\ref{H02}), which incorporates all cross-couplings of the fields into the transmon transitions. We characterize the pulse amplitude asymmetry of the Gaussian pulses by a parameter 
$\eta = (\Omega_{12} - \Omega_{01})/(\Omega_{12} + \Omega_{01})$ and we shift the counterdiabatic pulses 
by a quantity $\delta t_{02}$, $\Omega_{\rm 02} (t) \rightarrow \Omega_{\rm 02} (t-\delta t_{02})$. 
From  Fig. \ref{fig:calibrations} we observe the existence of a rather large plateau of transferred population around $\delta t_{02} =0$, showing a quite remarkable insensitivity to the STIRAP pulse symmetry. For the experiments, we choose to operate at two points, ($\eta = - 0.22$, $\delta t_{02} =0$) and ($\eta =-0.09$, $\delta t_{02} = 0$) which are somewhat in the middle of one of the plateaus and therefore are less exposed to errors.  Using these pulses we typically reach experimental values for $p_2$ in the range 0.8 - 0.9.

\section{Results}

\vspace{2mm}
\subsection{Synthetic gauge-invariant phase}

The calibration procedure described above was done by optimizing one of the phases of the three pulses while keeping the other two fixed. This is allowed by the gauge-invariance of the system with respect to the total circular phase $\Phi = \phi_{01} + \phi_{12} + \phi_{20}$ as we will demonstrate explicitly here.


To show this, we first examine the coupling Hamiltonian Eq. (\ref{eq:full_hamiltonian}) 
\begin{equation}
H(t) = \frac{\hbar}{2} \Omega_{01}(t) \hat{{\bf n}}_{01}\cdot {\bf \Lambda}_{01} + \frac{\hbar}{2} \Omega_{12}(t) \hat{{\bf n}}_{12}\cdot {\bf \Lambda}_{12}+ \frac{\hbar}{2} \Omega_{02}(t) \hat{{\bf n}}_{02}\cdot {\bf \Lambda}_{02},
\end{equation}
comprising the driving fields that couple into each pair of states $k,l \in \{0,1,2 \}$
with Rabi frequencies $\Omega_{jk}$ (real and positive) and phases $\phi_{jk}$. This describes three simultaneous rotations in the three subspaces $0-1$, $1-2$, and $0-2$ around the vectors $\hat{{\bf n}}_{kl}$. In each of the subspaces $(k,l)$, the action of the Hamiltonian is analogous to that of a spin-1/2 particle in a magnetic field of magnitude $\Omega_{12}$ and direction $\hat{{\bf n}}_{kl}$. For a single spin-1/2 particle it is always possible to rotate the axis so that one of them overlaps with the direction of the magnetic field. Crucially, for the three-level system it is not possible to  rotate arbitrarily all the three vectors $\hat{{\bf n}}_{kl}$. Indeed, by applying a unitary $U = e^{-i \chi_0} |0\rangle \langle 0| + e^{-i \chi_1} |1\rangle \langle 1|+ e^{-i \chi_2} |2\rangle \langle 2|$
one obtains a Hamiltonian $UHU^{\dagger}$ with a similar structure as Eq. (\ref{eq:full_hamiltonian}) but with different angles $\phi_{kl}'$; these new angles are not arbitrary, but they satisfy the constraint $\phi_{01}' + \phi_{12}' + \phi_{20}' = \phi_{01} + \phi_{12} + \phi_{20} = \Phi$. By performing local gauge transformations we can always eliminate two of the phases but the third one will be constrained by the value of the gauge-invariant quantity $\Phi$.  The situation is mathematically similar with that of a three-site system with complex hopping elements (Peierls hopping) \cite{Peierls:1933} and a magnetic field piercing the plaquette and creating a flux $\Phi$ \cite{Goldman:2014,Dalibard:2011}. 
This conclusion holds also for the total Hamiltonian $H + H_{\rm ac Stark}$, since by inspecting the ac Stark part Eq. (\ref{eq:acStark})
\begin{equation}
H_{\rm ac Stark}(t) = \frac{\hbar}{2} \sum_{j} \epsilon_{j}(t) |j\rangle \langle j| ,
\end{equation}
we have $UH_{\rm ac Stark}U^{\dag} = H_{\rm ac Stark}$. We can then define the Wilson loop around the triangle contour as 
\begin{equation}
W_{\triangle} = e^{i \phi_{01} + \phi_{12} + \phi_{20}} = e^{i \Phi},
\end{equation}
which is the path-ordered product of link variables $\exp(i \phi_{jk}) \in U(1)$.

\begin{figure}[tbp]
	\centering
	\includegraphics[width = 0.98\columnwidth]{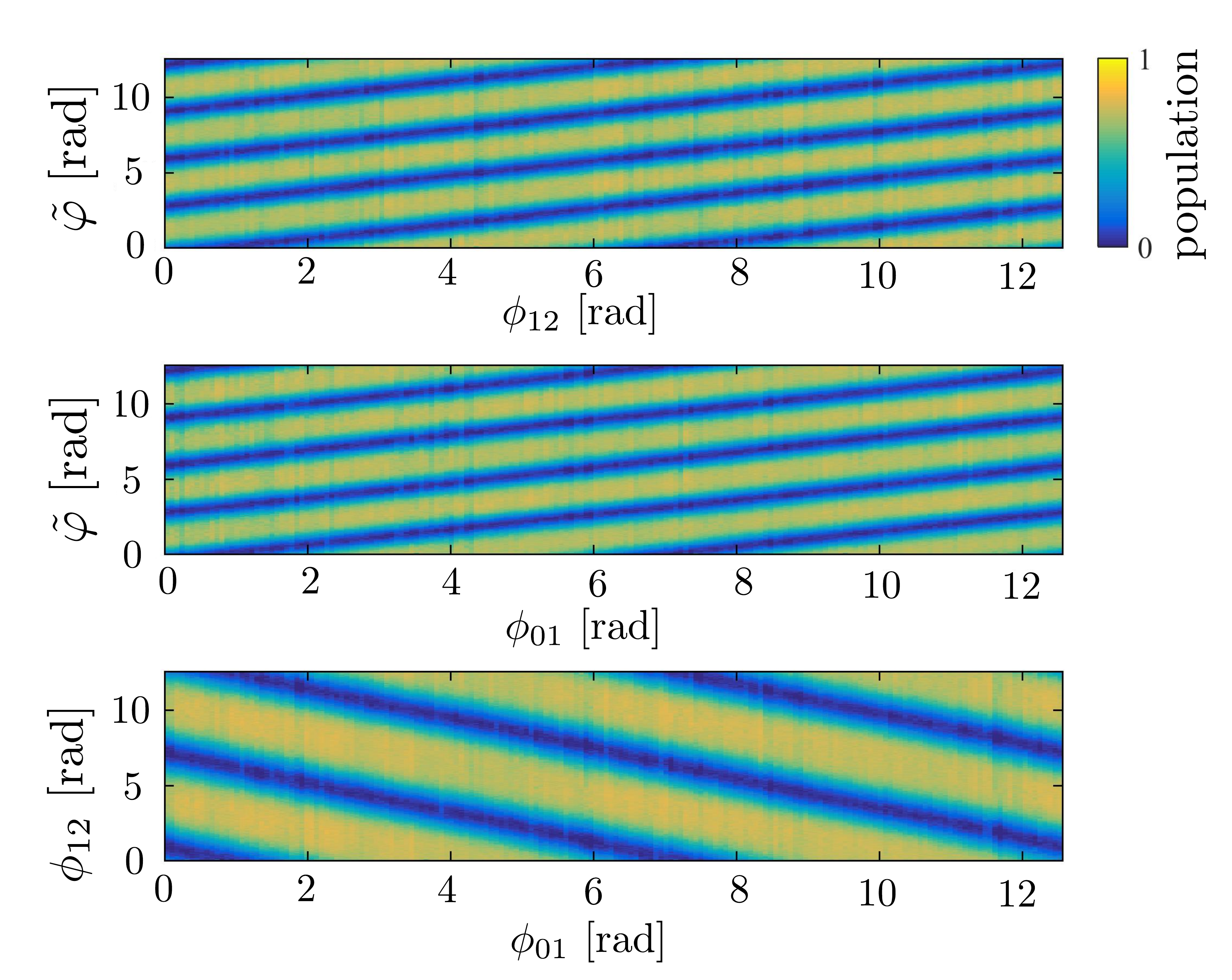}		
	\caption{Transferred population $p_{2}$ as a function of the phases $\phi_{01}$, $\phi_{12}$ and $\tilde{\varphi}$ of the externally-applied microwave fields. The three plots correspond, from the upper to the lower picture, to $\phi_{01}=0$, $\phi_{12}=0$ and $\tilde{\varphi}=0$ respectively; the other parameters for these measurements are $\Omega_{01}/(2\pi ) = 25$ MHz, $\Omega_{12}/(2 \pi ) = 16$ MHz ($\eta = -0.22$), $t_{\rm s} = -45$ ns,  and $\sigma = 30$ ns.
		The slope of the constant-population lines are positive in the two upper pictures and negative in the lowest one, and the $\tilde{\varphi}$ periodicity is twice as large as that of the periodicity in $\phi_{01}$ and  $\phi_{12}$. This verifies the relation $\phi_{01} +  \phi_{12} - 2\tilde{\varphi} - \pi = \Phi$. }
	\label{fig:saSTIRAP_full_measurement}
\end{figure}

In Fig. \ref{fig:saSTIRAP_full_measurement} we present the population transferred to state $\ket{2}$ using saSTIRAP when either of the angles $\phi_{01}$, $\phi_{12}$ and $\tilde{\varphi}$ are varied, while keeping the other two fixed. The populations are measured at a time $t=20$ ns. 
The experiment shows clearly that the method can successfully transfer population to state $\ket{2}$, given the correct choice of the phases and shows that the three phases for a given transferred population are not independent from each other.
From the data, the $\pi$-periodicity of the population transferred as a function of the phase $\tilde{\varphi}$ of the two-photon drive pulse is also manifest. In contradistinction, a sequential process (where we populate the first excited state, then transfer to the second excited state) should display a $2 \pi$ periodicity in the single-photon drive phase. This demonstrates the fully quantum-coherent nature of the process.

Once the phenomenon of gauge invariance is demonstrated, we can proceed by fixing the gauge. A convenient choice is $\phi_{01}' = 0$, $\phi_{12}' = 0$, and $\phi_{20}'= \Phi$, which leads to the following structure for Eq. (\ref{eq:full_hamiltonian}),
\begin{equation}
H_{\Phi}(t) = \frac{\hbar}{2} \Omega_{01}(t) \Lambda_{01}^{s} + \frac{\hbar}{2} \Omega_{12}(t) \Lambda_{12}^{s}+ \frac{\hbar}{2} \Omega_{02}(t) \hat{{\bf n}}_{\Phi}\cdot {\bf \Lambda}_{02}, \label{eq:full_hamiltonian_standard}
\end{equation}
where $\hat{{\bf n}}_{\Phi} = (\cos \Phi , \sin \Phi )$. This form puts in evidence the role of the gauge-invariant phase $\Phi$ as a parameter in the Hamiltonian, which can be controlled experimentally along with the Rabi frequencies $\Omega_{01}(t)$, $\Omega_{12}(t)$, and $\Omega_{02}(t)$.

\subsection{Broken time-reversal symmetry}

For a spin lattice the time-reversal symmetry is relatively straightforward to understand. A magnetic field, either applied externally or resulting from the complex phases of the couplings, remains invariant when time runs backwards; as a result, the time-reversed Schr\"odinger equation is no longer satisfied. This leads to non-reciprocal phenomena: an input signal at one lattice site might be transmitted to another site, but nothing will be transmitted if we reverse the direction of the signal. In practice, this can be used for realizing non-reciprocal devices such as circulators or isolators.

Here we examine in detail the time-reversal symmetry of the problem. We first note that STIRAP itself is clearly time-symmetric. Indeed, starting from state $|2\rangle$ as the initial state and running backwards in time the STIRAP process, the system will see the $\Omega_{01}$ wave as the Stokes pulse and then the $\Omega_{12}$ wave as the pump pulse, thus realizing the usual counterintuitive sequence. Experimentally, STIRAP reversal has been demonstrated already in  \cite{stirap_ours}. This can be understood by recalling that in this situation the wavefunction simply follows the slow variation of the Hamiltonian dark state as the mixing angle varies from  $0$ to $\pi /2$. Thus, when reversing the direction of time, $\Theta \rightarrow \pi /2 - \Theta$ and $\cos \Theta |0\rangle - \sin \Theta |2\rangle \rightarrow - \left(\cos \Theta |2\rangle - \sin \Theta |0\rangle \right)$, with the roles of the states $|2\rangle$ and $|0\rangle$ reversed, as expected. 

The situation changes dramatically when the additional transfer path provided by the two-photon pulse is introduced and the gauge-invariant phase $\Phi$ is established. In the spin system one sees immediately that this is equivalent to the appearance of a magnetic field piercing the plaquette. We do expect then to have a broken time symmetry if this magnetic field is non-zero, and a time-symmetric problem otherwise, and similar considerations will hold for the three-level system.


\begin{figure*}[tbp]
	\centering
	\includegraphics[width = 1.0\textwidth]{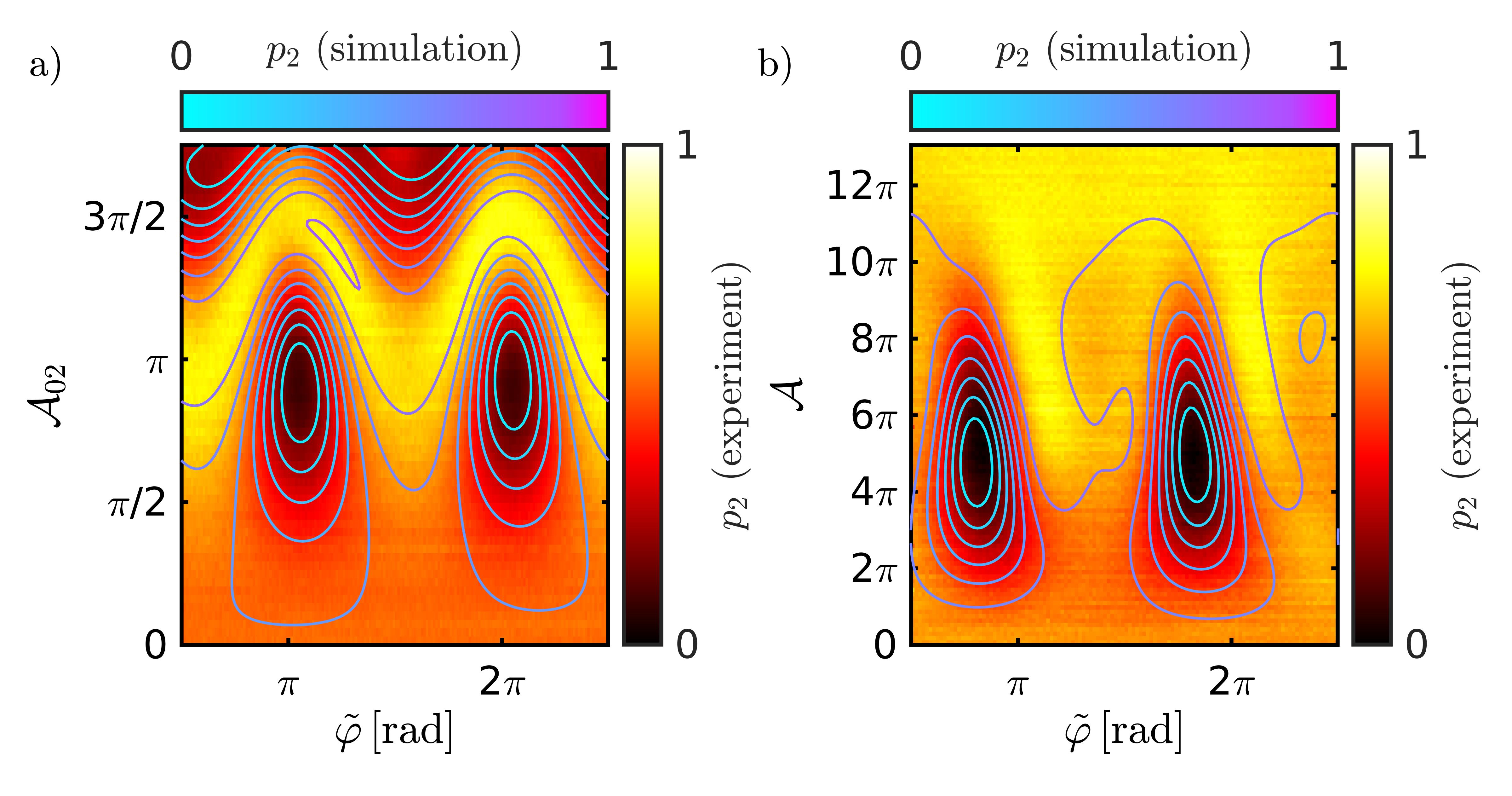}
	\caption{Observation of broken time-reversal symmetry, with experimental data shown as a continuum and the simulations as contour plots. We measure the transferred population as a function of the gauge-invariant phase $\Phi$ (controlled experimentally via the phase $\tilde{\varphi}$) at increasing values of the a) two-photon pulse area $\mathcal{A}_{02}$ and b) STIRAP area $\mathcal{A}$. The experimental parameters were a) $t_{\rm s} = -61$ ns, $\sigma = 25$ ns, $\Omega_{01}/(2 \pi ) = 44$ MHz, $\Omega_{12}/(2 \pi )  = 36.8$ MHz, $\mathcal{A} = 5.5\pi$ and b) $t_{\rm s} = - 37.5$ ns  $\sigma = 25$ ns, and
		$\mathcal{A}_{02} = 0.81 \pi$. The transition into the regime of broken time-reversal symmetry (corresponding to non-reciprocality in the spin system) is seen through the gradual emergence of $\Phi$-dependence as a) $\mathcal{A}_{02}$ or b) $\mathcal{A}$ is increased.
	}
	\label{fig:areaandphase}
\end{figure*}

As usual in time-reversal problems, we define an antilinear complex conjugation operator ${\cal K}$; when applied from the right to the Schr\"odinger equation 
\begin{equation}
i \hbar \frac{\partial |\psi (t)\rangle}{\partial t} = H_{\Phi} (t) |\psi (t) \rangle  \label{eq:beforereversal}
\end{equation}
we obtain
\begin{equation}
-i \hbar \frac{\partial}{\partial t} {\cal K}|\psi (t) \rangle =  {\cal K}H_{\Phi} (t){\cal K} {\cal K}|\psi (t) \rangle  ,\label{eq:afterreversal}
\end{equation} 
where we used ${\cal K}^2 =1$. 
A time-reversed Schr\"odinger equation
\begin{equation}
i \hbar \frac{\partial |\psi' (t')\rangle}{\partial t'} = H'_{\Phi'} (t') |\psi' (t') \rangle , 
\end{equation}
where $t' = -t$, can be obtained by identifying $|\psi' (t') \rangle = {\cal K}|\psi (t) \rangle$ and $H'_{\Phi'} (t') = {\cal K}H_{\Phi} (t){\cal K}$.
By examining Eq. (\ref{eq:full_hamiltonian_standard}) we notice that ${\cal K}H_{\Phi} (t){\cal K}= H_{-\Phi + 2n\pi} (t)$. Thus, the time-reversed evolution corresponds to changing the gauge-invariant phase to $\Phi'=-\Phi + 2n\pi$ (or $\tilde{\varphi}'= - \tilde{\varphi} - (n+1)\pi$), and  the time-reversal symmetry is broken for all values of $\Phi$, with the exception of $\Phi = n\pi$ (or $\tilde{\varphi} = - (n+1) \pi /2$). These results agree also with  the findings in \cite{Koch2010} for a three-site lattice, corresponding to reversing the direction of magnetic field piercing the lattice in Fig. 1 b). It is important to understand that these considerations do not depend on the particular gauge used in Eq. (\ref{eq:full_hamiltonian_standard}): the same conclusion is reached if the Hamiltonian Eq. (\ref{eq:full_hamiltonian}) is examined. This is due to the fact that no gauge transformation can make the Hamiltonian Eq. (\ref{eq:full_hamiltonian}) real, with the exception of the case $\Phi = \phi_{01} + \phi_{12} + \phi_{20}= n\pi$. In the spin lattice, this case corresponds to an integer number of flux quanta per unit cell. Also, because the Hamiltonian $H_{\rm acStark}$ is invariant under time-reversal (there is no phase-dependence in the ac Stark shifts), the breaking of the time-reversal symmetry due to $H_{\Phi}$ should be observable when the system is evolved under the full Hamiltonian comprising also the inhomogenous/ac-Shifted part.

To demonstrate the gradual onset of the broken time-symmetry regime,  we perform the experiments shown in Fig. \ref{fig:areaandphase}, where we measure the transferred populations at different phases $\tilde{\varphi}$. 
We introduce the area of the counterdiabatic pulse $\mathcal{A}_{02} = \int_{-\infty}^{\infty} {\mathrm d}t\, \Omega_{02}(t)$
and we define the STIRAP pulse area as $\mathcal{A} = \int_{-\infty}^{\infty} {\mathrm d}t\,\sqrt{\Omega_{01}^{2}(t) + \Omega_{12}^{2}(t)}$ 
which is a measure of adiabaticity of the STIRAP process. In Fig. \ref{fig:areaandphase}a) we show the population of state $\ket{2}$ as a function of the area of the counterdiabatic pulse and phase.  We can also examine the dependence of the population $p_2$ on the STIRAP area $\mathcal{A}$, while keeping the counterdiabatic pulse area $\mathcal{A}_{02}$ constant, see  
\ref{fig:areaandphase}b). 
As expected from the previous gauge-invariance considerations and the use of a two-photon transition $|0\rangle \rightarrow |2\rangle$, the transfer is $\pi$-periodic in $\tilde{\varphi}$. One notices however small deviations from perfect $\pi$-periodicity especially in Fig. \ref{fig:areaandphase}b), reflecting the limitations of the two-photon approximation. 

We further observe the main features of broken time-reversal symmetry:
 when both the STIRAP process and the counteradibatic fields are on, the transfer is in general not symmetric under the transformation  $\tilde{\varphi} \rightarrow - \tilde{\varphi} - (n+1)\pi$. The plots also show that if either one of the couplings is turned off, time-symmetry is restored. For example, in Fig. \ref{fig:areaandphase}a) there is no phase dependence for $\mathcal{A}_{02} = 0$. Similarly, from  Fig. \ref{fig:areaandphase}b) we notice the absence of phase dependence for $\mathcal{A} = 0$, as expected when only the two-photon pulse is applied, while in the other extreme case, at large values
$\mathcal{A} > 12 \pi$, STIRAP dominates and the phase dependence becomes again weaker. 
In general, these patterns of transfer are complicated but they are reproduced very well by the numerical modeling of the total Hamiltonian (contour plots). For the ideal case of unitary evolution under $H_{\Phi}(t)$, the maximum transfer should occur at $\mathcal{A}_{02} = \pi$ (which follows immediately by using $\Omega_{02}(t) = 2\dot{\Theta}(t)$) and at $-2\tilde{\varphi} - \pi = \Phi= -\pi/2 + 2n\pi$ (or $\tilde{\varphi}\approx 3 \pi/4 + n\pi$ ). However, in general the presence of the Hamiltonian $H_{\rm acStark}$ shifts these optimal phases and areas to other values (see Supporting Information  SI2), depending on the specific parameters of the pulses. Still, the values of the measured optimal phases are reproduced quite well by the numerics, typically within $0.09\pi$. In these regions of optimal transfer, the population on state $|2\rangle$ exceeds 0.9. 

This experiment also shows, as emphasized in \cite{Koch2010}, that an intimate connection exists between gauge invariance and time symmetry, which experimentally amounts to the fact that time-reversal symmetry is fully controlled by only one parameter, the gauge-invariant phase $\Phi$.

\subsection{Currents and chirality}

For transport phenomena in spin chains, the measurement and analysis of the currents provide important insights into the dynamics. Much attention has been given so far to the case where the currents exibit a circular flow, which can be made clockwise or anticlockwise by changes in the gauge-invariant phase \cite{Roushan2017,Scully2019}. Here we will show that the analysis of currents give important insights into the mechanism by which the number of excitations transferred from one site to another is maximized by adiabatic processes. 

To start with, let us calculate the time-dependent currents in a dark state. Using Eq. (\ref{currents}) we find
\begin{equation} 
\langle D (t)|I_{0\rightarrow 2}| D(t) \rangle = \frac{1}{2}\Omega_{02} (t) \sin 2 \Theta (t) \sin \phi_{02}, \label{current02}
\end{equation}
and 
\begin{equation} 
\langle D (t)|I_{0\rightarrow 1}| D(t) \rangle = \langle D (t)|I_{1\rightarrow 2}| D(t) \rangle = 0 \label{current01and12}.
\end{equation}

We notice that current $I_{0\rightarrow 2}$ depends not only on $\Omega_{02}$, as expected, but also on the mixing angle $\Theta (t)$. Clearly for equal-strenght STIRAP pulses $\Omega_{01} = \Omega_{12}$ the maximum current is realized in the middle of the protocol, that is at $t=t_{s}/2$. 
The magnitude of the current is modulated by the $\sin \phi_{02}$ factor. For $\phi_{02} = \pi /2$ (or $\Phi = -\pi /2$) we obtain a maximum current 
\begin{equation} 
\langle D (t)|I_{0\rightarrow 2}^{\rm max}| D(t) \rangle = \frac{1}{2}\Omega_{02} (t) \sin 2 \Theta (t), \label{current02max}
\end{equation}
in the direction of increasing the population on the state $|2\rangle$. For $\phi_{02} = -\pi /2$ (or $\Phi = \pi /2$) the current would flow in the opposite direction: the transfer realized by STIRAP is undone by the two-photon pulse. The fact that the averages of $I_{0\rightarrow 1}$ and $I_{0\rightarrow 1}$ on the dark state are zero reflects the fact that the state $|1\rangle$ is not populated. The result has a paradoxical flavor, since a quanta is transferred along a trajectory without apparently going through the intermediate positions, which leads to infinite Bohmian velocities at those positions \cite{Benseny2012}. 

Let us examine now  the time-derivative of the population $p_{2}(t) = \sin^{2} \Theta (t)$; we find
\begin{equation}
\dot{p}_{2}(t)= \dot{\Theta}(t)\sin 2\Theta (t) .\label{current:STIRAP}
\end{equation}

We can immediately compare Eqs. (\ref{current02}, \ref{current01and12}) with Eq. (\ref{current:STIRAP}). To have consistency between these results, we need to impose the condition $\Omega_{02} (t) = 2 \dot{\Theta}(t)$ and $\phi_{02} = \pi /2$. These are precisely the requirements of superadiabatic driving.

\begin{figure}[tbp]
	\centering
	\includegraphics[width = 0.98\columnwidth]{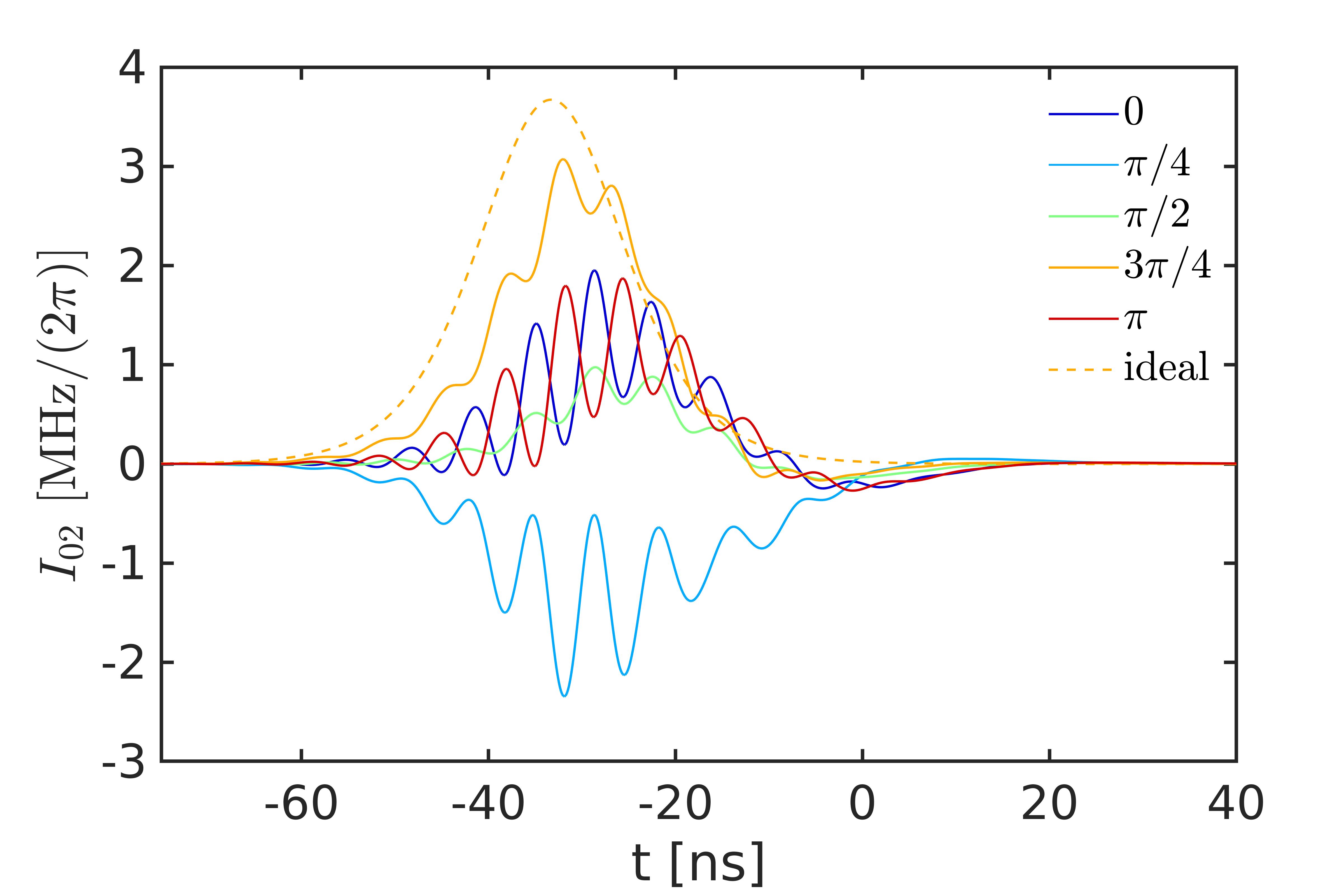}	
	\caption{The averaged 0 - 2  current $\langle I_{02} \rangle$ at different values of $\tilde{\varphi}$. The parameters are the same as in Fig. 5a) for a counterdiabatic pulse area $\mathcal{A}_{02} =\pi$.
	}
	\label{fig:current02}
\end{figure}

Next, we give a more precise account of the intuition that saSTIRAP can be seen as a constructive interference effect between two paths, one corresponding to the STIRAP process and the other to the two-photon process. This argument illustrates surprisingly well why a purely imaginary value for the $0 - 2$ driving is necessary.

Let us consider the case ${\cal A}=0$. Since the state $|1\rangle$ is not populated, let us consider the two-photon process with Hamiltonian
\begin{equation}
H_{02} = \frac{\hbar}{2} \Omega_{02} \hat{\bf n}_{\Phi}\cdot{\bf \Lambda}_{02},
\end{equation}
where $\hat{\bf n}_{\Phi} = (\cos \Phi , \sin \Phi )$ is a unit vector in the plane $xOy$ and ${\bf \Lambda}_{02} = (\Lambda_{02}^{s},\Lambda_{02}^{a})$ is the vector formed by the symmetric and antisymmetric Gell-Mann matrices.
In this subspace the evolution operator is
\begin{equation}
U(t) = e^{-\frac{i}{\hbar}\int_{-\infty}^{t}d\tau H_{02}(\tau )}= e^{-\frac{i}{2}\int_{-\infty}^{t}d\tau \Omega_{02}(\tau ) \hat{\bf n}_{\Phi} {\bf \Lambda}_{02}}.
\end{equation}
Now, the components of ${\bf \Lambda}_{02}$ are essentially Pauli matrices in the $0-2$ subspace (and all the other matrix elements are zero), therefore we can use a familiar formula for the exponential of Pauli matrices,
\begin{equation}
e^{i {\bf n}_{\Phi}\cdot {\bf \Lambda}_{02}} = \cos n + i \hat{\bf n}_{\Phi}\cdot{\bf \Lambda}_{02} \sin n,
\end{equation}
with $n = -\frac{1}{2}\int_{-\infty}^{t}d\tau \Omega_{02}(\tau )$ and ${\bf n}_{\Phi} = n \hat{\bf n}_{\Phi}$, resulting in
\begin{eqnarray}
U(t) &=& \cos \left( \frac{1}{2} \int_{-\infty}^{t}d\tau \Omega_{02}(\tau ) \right) \nonumber \\ 
& & - i \hat{\bf n}_{\Phi} \cdot{\bf \Lambda}_{02} \sin \left( \frac{1}{2} \int_{-\infty}^{t}d\tau \Omega_{02}(\tau ) \right).
\end{eqnarray}
Since $\Omega_{02}(t) = 2 \dot{\Theta} (t)$, the state at any time is obtained as
\begin{equation}
|0\rangle \rightarrow U(t)|0\rangle = \cos \Theta (t) |0\rangle + e^{i (\Phi - \pi /2)} \sin \Theta (t) |2\rangle
\end{equation}
It is instructive to see that this state coincides with the dark state precisely for $\Phi = -\pi /2$ as expected.

Now, for a $\pi$ pulse from $t=-\infty$ to $t=\infty$ we have $\int_{-\infty}^{\infty}d\tau \Omega_{02}(\tau ) = \pi$, resulting in
\begin{equation}
U_{\pi } = -i\left(\cos \Phi \Lambda_{02}^{s} + \sin \Phi \Lambda_{02}^{a} \right).
\end{equation}
When applied to the initial state $|0\rangle$, this leads to
\begin{equation}
U_{\pi} |0\rangle = \left( -i \cos \Phi + \sin \Phi \right)|2\rangle .
\end{equation}
Suppose now that $\Phi = -\pi/2$ (up to integer multiples of $2\pi$). This means that $U_{\pi} |0\rangle = -|2\rangle$.  
The same sign is obtained from the STIRAP path, $|0\rangle \stackrel{\rm STIRAP}{\longrightarrow} -|2\rangle$, therefore we expect that these paths will interfere constructively.
Conversely, if $\Phi = \pi/2$ (up to integer multiples of $2\pi$), we expect destructive interference, since $U_{\pi} |0\rangle = +|2\rangle$. This is precisely what is observed in the experiment.
That is, the dynamics along the STIRAP path occurs in the $\{|0\rangle , |2\rangle \}$ subspace.

In the experiments with the three-level simulator, the currents can be obtained by calculating the averages of the operators Eq. (\ref{currents})
on the state extracted from experimental data. In Fig. \ref{fig:current02} we present the current  $\langle I_{02} \rangle$ at a few values of $\tilde{\varphi}$ for $\mathcal{A}_{02} =\pi$ and with the rest of the parameters as in Fig. 5a). In general, the features we observe are consistent with the idealized model above; in addition, oscillations are present in the currents due to the ac Stark shift. At  $\tilde{\varphi}= 3\pi/4 + n \pi$ we obtain a relatively large positive current. The envelope of this current matches well with the ideal-case analytical expression Eq. (\ref{current02max}), plotted with dotted line. As we depart from this optimal transfer point, the current becomes more oscillatory and smaller in value. It can even have negative values for points in the regions of minimal population transfer, as shown in Fig. \ref{fig:current02} for $\tilde{\varphi}= \pi/4$, signaling the transfer of population backwards to state $|0\rangle$. Note also that the points $-\pi/4$ and $3\pi/4$ are related by the time-reversal relation $\tilde{\varphi}' = -\tilde{\varphi} - (n+1)\pi$; thus, as expected, the currents show conclusively the signature of broken time-reversal symmetry. Finally, let us notice that the dark state in STIRAP and saSTIRAP involves superpositions of states with various chiralities. However, the average of the chirality operator on this state is zero,
\begin{equation}
\langle D(t)| C_{z} |D(t)\rangle =0,
\end{equation}
reflecting the connection between chirality and asymmetry, namely the fact that the chirality is expressed in terms of only asymmetric Gell-Mann matrices Eq. (\ref{eq:chiralityGellMann}). In this sense, increasing ${\cal A}$ in Fig. \ref{fig:areaandphase} b) also results in a change of chirality. For ${\cal A}=0$ we have 
\begin{equation}
\langle 0 | U^{\dag} (t) C_{z} U (t)|0\rangle = 
\frac{\sqrt{3}}{3} \cos \Phi \sin 2 \Theta (t), 
\end{equation}
which is zero only at the beginning and at the end of the process ($\Theta = 0, \pi /2$ ) if $\Phi \neq \pm \pi/2$. 

This is easy to understand in the Bloch-sphere picture for the subspace $\{|0\rangle , |2\rangle \}$. There, the dark state moves from the North pole to the South pole in the $z-O-x$ plane, while the chirality becomes the Pauli-y operator. The average 
value of the $y$-axis projection will be therefore zero at any time for the dark state.

\section{Conclusions}
We have employed a transmon superconducting circuit in the loop driving ($\Delta$-driving) configuration as a simulator for a spin chain with XX and Dzyaloshinskii-Moryia couplings and subjected to time-dependent inhomogenous broadening.  We demonstrate that transport can be realized efficiently under the condition of superadiabaticity. We put in evidence the phenomenon of gauge invariance and we observe the manifestation of broken time-reversal symmetry. Finally, we extract the currents and show that the superadiabaticity condition leads to a maximum positive current flowing between the initial state and the target state. 

\newpage

\hspace{3mm} {\bf\Large Supporting Information}

\section*{SI1: Driven three-level system in the Gell-Mann representation}

\subsection{Notations} 

The Gell-Mann matrices offers a compact representation of the Hamiltonian of the three-level transmon simulator, which is further amenable to straightforward manipulations based on symmetries.

To recall, the Gell-Mann matrices are  $3\times 3$ matrices which form a representation of the 8 generators of the Lie algebra $su(3)$ associated with the group SU(3). They are traditionally denoted in quantum chromodynamics by ${\lambda_1, ..., \lambda_8}$.
As the Gell-Mann matrices are a direct generalization of the Pauli matrices for higher dimensions, we will use here a notation that puts in evidence precisely this feature. Firstly, we will work in a rotating frame defined by the three pairs of states, and each of these pairs is coupled resonantly by single-photon transitions ($|0\rangle \leftrightarrow|1\rangle$ and $|1\rangle \leftrightarrow |2\rangle$) and by the two-photon transition ($|0\rangle \leftrightarrow |2\rangle$). As a result, the matrices $\lambda_3$ and $\lambda_8$ do not appear in the Hamiltonian Eq. (\ref{eq:Hamiltonian_couplings}), since they contain diagonal elements. The remaining 6 matrices are off-diagonal, and can be classified as symmetric (in analogy with the $\sigma^x$ Pauli matrix) and antisymmetric (in analogy with the $\sigma^y$ Pauli matrix). Specifically, we define
\begin{eqnarray}
\Lambda_{01}^{s} \equiv \lambda_1 =  \left[\begin{array}{ccc} 0 & 1 & 0 \\ 1 & 0 & 0 \\ 0 & 0 & 0 \end{array} \right], & ~\Lambda_{01}^{a} \equiv \lambda_2 =  \left[\begin{array}{ccc} 0 & -i & 0 \\ i & 0 & 0 \\ 0 & 0 & 0 \end{array} \right], \nonumber \\
\Lambda_{12}^{s} \equiv \lambda_6 =  \left[\begin{array}{ccc} 0 & 0 & 0 \\ 0 & 0 & 1 \\ 0 & 1 & 0 \end{array} \right], & ~\Lambda_{12}^{a} \equiv  \lambda_7 =  \left[\begin{array}{ccc} 0 & 0 & 0 \\ 0 & 0 & -i \\ 0 & i & 0 \end{array} \right], \nonumber \\
\Lambda_{02}^{s} \equiv  \lambda_4 =  \left[\begin{array}{ccc} 0 & 0 & 1 \\ 0 & 0 & 0 \\ 1 & 0 & 0 \end{array} \right], & ~\Lambda_{02}^{a} \equiv \lambda_5 =  \left[\begin{array}{ccc} 0 & 0 & -i \\ 0 & 0 & 0 \\ i & 0 & 0 \end{array} \right]. \nonumber 
\end{eqnarray}

In terms of Fock states these read $\Lambda_{kl}^{a} = - \Lambda_{lk}^{a} =  -i|k\rangle\langle l| + i|l\rangle\langle k|$ and $\Lambda_{kl}^{s} =  \Lambda_{lk}^{s} = |k\rangle \langle l| +  |l\rangle \langle k|$, and the symmetry/asymmetry property can be written as $\Lambda_{ij}^{a} = - \Lambda_{ji}^{a}$ and $\Lambda_{ij}^{s} =  \Lambda_{ji}^{s}$.

\subsection{Effective Hamiltonian in the loop driving configuration}

To obtain the representation Eq. (\ref{eq:Hamiltonian_couplings}), we employ an interaction picture with respect to the undriven Hamiltonian by applying the transformation $U_{\rm I}(t)= |0\rangle\langle 0| + \exp (i \omega_{01}t)|1\rangle \langle 1| + \exp [i (\omega_{01} + \omega_{12})t] |2\rangle \langle 2|$, resulting in $\mathcal{H} \rightarrow U_{\rm I} \mathcal{H} U_{\rm I}^{\dag} + i\hbar (\partial_{t}U_{\rm I}) U_{\rm I}^{\dag}$. 
We separate the total Hamiltonian $\mathcal{H}$ resulting after this transformation into a part $\mathcal{H}_{01} + \mathcal{H}_{12}$ that corresponds to couplings via the fields $\Omega_{01}$ and 
$\Omega_{12}$ used in STIRAP, and a part $\mathcal{H}_{02}$ produced by two-photon driving.

For the STIRAP part we have
\begin{eqnarray}
&\mathcal{H}_{01}(t)& = \hbar\left[\Omega_{01}(t) \cos(\omega_{01}t + \phi_{01}) + \right. \nonumber  \\ 
&&\left. + \frac{\Omega_{12}(t)}{\sqrt{2}}\cos(\omega_{12}t + \phi_{12}) \right]e^{-i \omega_{01} t} |0\rangle \langle 1| + {\rm h.c.}  \label{H01} \\
&\mathcal{H}_{12}(t)&= \hbar\left[\sqrt{2}\Omega_{01}(t) \cos(\omega_{01}t + \phi_{01}) +\right. \nonumber  \\
&& \left. + \Omega_{12}(t) \cos(\omega_{12}t + \phi_{12})\right] e^{-i \omega_{12} t} |1\rangle \langle 2|+ {\rm h.c.} \label{H12}
\end{eqnarray}
Here the factors of $\sqrt{2}$ in the cross-coupling terms are due to the increase by $\sqrt{2}$ of the matrix elements as we go from the first to the second transition. In the rotating wave approximation, by neglecting terms oscillating at frequencies $\omega_{01}\pm \omega_{12}, 2\omega_{12}, 2\omega_{01}$, we find 
\begin{eqnarray}
H_{01}(t) &=& \frac{\hbar}{2}\Omega_{01}(t) e^{i \phi_{01}} |0\rangle \langle 1| + h.c. \\
&=& \frac{\hbar}{2} \Omega_{01} \left( \cos\phi_{01} \Lambda_{01}^{s} - \sin\phi_{01} \Lambda_{01}^{a} \right) , \label{eq:H01}\\
H_{12} (t) &=& \frac{\hbar}{2}\Omega_{12}(t) e^{i \phi_{12}} |1\rangle \langle 2| + h.c. \\ 
&=& \frac{\hbar}{2} \Omega_{12} \left( \cos\phi_{12} \Lambda_{12}^{s} - \sin\phi_{12} \Lambda_{12}^{a} \right). \label{eq:H12}
\end{eqnarray}

To drive the $|0\rangle \leftrightarrow |2\rangle $ transition we use a single microwave field with frequency $\tilde{\omega}$ and phase $\tilde{\varphi}$, such that the two-photon resonance condition $2\tilde{\omega}= \omega_{01}+\omega_{12}$ holds. This tone is detuned from the $0 - 1$ and $1 - 2$ transitions by $\Delta = \tilde{\omega} - \omega_{12} = \omega_{01} -  \tilde{\omega} = (\omega_{01} - \omega_{12})/2$. We denote by $\tilde{\Omega}_{01}$ and $\tilde{\Omega}_{12}$ the Rabi couplings of this field into the $0 - 1$ and $1 -2$ transitions respectively, noting again that in the weak anharmonicity approximation for the transmon
$\tilde{\Omega}_{12} \approx \sqrt{2} \tilde{\Omega}_{01}$.
The two-photon field results in the Hamiltonian
\begin{eqnarray}
\mathcal{H}_{02}(t) &=& \hbar\tilde{\Omega}_{01}(t)\cos (\tilde{\omega}t + \tilde{\varphi}) e^{-i\omega_{01}t}|0\rangle \langle 1| + \nonumber \\ 
& & + \hbar\tilde{\Omega}_{12}(t)\cos (\tilde{\omega}t + \tilde{\varphi})  e^{-i\omega_{12}t}|1\rangle \langle 2| + h.c. \label{H02}
\end{eqnarray}
We neglect the fast rotating terms at $\tilde{\omega} + \omega_{12}$ and $\tilde{\omega} + \omega_{01}$ and obtain
\begin{eqnarray}
\tilde{H}_{\rm 2ph}(t) &=& \frac{\hbar}{2}\left[\tilde{\Omega}_{01}(t)e^{-i\Delta t + i\tilde{\varphi}}|0\rangle \langle 1| + \tilde{\Omega}_{12}(t)e^{+i\Delta t + i \tilde{\varphi}}|1\rangle \langle 2 |\right] \nonumber \\ 
& & + h.c.
\end{eqnarray}
which produces \cite{James2007} a two-photon complex coupling with Rabi frequency $\Omega_{02} = \tilde{\Omega}_{01}\tilde{\Omega}_{12}/(2\Delta)$ and 
phase $\phi_{02} = 2\tilde{\varphi}  + \pi$,
\begin{eqnarray}
H_{02} &=& -\frac{\hbar \tilde{\Omega}_{01}\tilde{\Omega}_{12}}{4 \Delta} e^{2i\tilde{\varphi}}|0\rangle \langle 2| + h.c. \\
&=& \frac{\hbar}{2} \Omega_{02} \left( \cos\phi_{02} \Lambda_{02}^{s} - \sin\phi_{02} \Lambda_{02}^{a} \right),
\end{eqnarray}
allowing us eventually to use this coupling as a counterdiabatic Hamiltonian $H_{\rm cd}(t)$ as in Eq. \eqref{eq:cd}. 
Note that the relative phase $2 \tilde{\varphi}$ between the counterdiabatic two-photon pulse and the STIRAP pulses is fixed during the evolution: once defined at one time, it will remain the same at any other time due to the frequency matching relation $2\tilde{\omega} = \omega_{01} + \omega_{02}$.


\subsection{Derivation of the counteradiabatic drive by the method of adiabatic potentials}

The standard way of deriving the exact form of the effective Rabi drive $\Omega_{02}$ is by applying the concept of reversed Hamiltonian enineering to the STIRAP case, see {\it e.g.}  \cite{Chen10,Giannelli14}. Another very elegant  method is by the introduction of adiabatic potentials \cite{Polkovnikov2017}, where the counterdiabatic Hamiltonian is identified by the contribution that would cancel excitations in a frame which follows the adiabatic change, see also \cite{Fleischhauer99,Unanyan97}. The adiabatic potential can be expressed in terms of commutators of the Hamiltonian and its derivatives with respect to the adiabatic control parameter \cite{Claeys2019}, and since the Gell-Mann matrices 
$\Lambda_{01}^{s}$, $\Lambda_{12}^{s}$, and	$\Lambda_{02}^{a}$ form a closed subalgebra (see also Eq. (\ref{algebra}) below), it should be automatically realizable provided that the Hamiltonians $H_{01}, H_{12}$, and $H_{02}$ are available. Following the general formalism \cite{Polkovnikov2017}, we identify the adiabatic control parameter as the STIRAP mixing angle $\Theta$, and therefore we have $H_{\rm cd}(t) = \dot{\Theta} \mathcal{A}_{\Theta}$, where $\mathcal{A}_{\Theta}$ is the adiabatic gauge potential. In the case of STIRAP, this potential can be calculated by analyzing the three instantaneous eigenvalues $|n_{\pm}\rangle$, $|n_{0}\rangle$ of the Hamiltonian with two driving fields $H_{01} + H_{12}$ derived in Eqs. (\ref{eq:H01},\ref{eq:H12}), which can be written in the convenient form \cite{photonics2016} 
\begin{eqnarray}
|n_{0}\rangle &=& |D\rangle ,\\
|n_{\pm} \rangle &=&  \frac{1}{\sqrt{2}}|B\rangle \pm \frac{e^{-i \Phi_{01}}}{\sqrt{2}}|1\rangle ,
\end{eqnarray}
where $|B\rangle = \sin\Theta |0\rangle + e^{-i (\phi_{01} + \phi_{12})} \cos \Theta |2\rangle$ is the bright state and  $|D \rangle = \cos\Theta |0\rangle - e^{-i (\phi_{01} + \phi_{12})} \sin \Theta |2\rangle$ is the dark state. The matrix elements of $\mathcal{A}_{\Theta}$ in this basis are 
obtained as $\langle m (\Theta) |\mathcal{A}_{\Theta} | n (\Theta) \rangle = i \hbar \langle m (\Theta)  |\partial_{\Theta} n (\Theta) \rangle$, where $|n (\Theta) \rangle$, $|m (\Theta) \rangle$ are any of the instantaneous eigenvalues $|n_{\pm}\rangle$, $|n_{0}\rangle$ obtained above. By noticing that $\partial_{\Theta} |D\rangle = - |B\rangle$ and $\partial_{\Theta} |B\rangle = |D\rangle$, we find 
\begin{equation}
\mathcal{A}_{\Theta} = i \hbar e^{i (\Phi_{01} + \Phi_{12})}|0\rangle \langle 2| -  i \hbar e^{-i (\Phi_{01} + \Phi_{12})}|2\rangle \langle 0|  .
\end{equation}
This allows us to identify the terms in the counterdiabatic Hamiltonian as $\Omega_{02} = 2 \dot{\Theta}$ and $\phi_{02} = \phi_{01} + \phi_{12} + \pi/2$, in agreement with the standard result \cite{Chen10,Giannelli14,Fleischhauer99,Unanyan97}.

\subsection{Spin-1 representation}

An interesting intuitive picture of saSTIRAP is obtained by further examining the properties of the relevant Gell-Mann matrices. Indeed, consider the case $\phi_{01}=\phi_{12}=0$ and $\phi_{02} = \pi/2$, which is realized in saSTIRAP. Then, the only nonzero terms in the total Hamiltonian are those containing the matrices $K_{x} \equiv \Lambda_{01}^{s}\equiv \lambda_{1}$, $K_{y} \equiv \Lambda_{12}^{s} = \lambda_{6}$,  
and	$K_{z} \equiv \Lambda_{02}^{a} = \lambda_{5}$. These matrices -- denoted now for  clarity $K_{j}$ with the index $j\in \{ x,y.z \}$ -- form a representation of the spin-1 angular-momentum operators, satisfying the commutation relations
\begin{equation}
[K_{i}, K_{j}] = i\epsilon_{ijk} K_{k}. \label{algebra}
\end{equation}
These matrices are traceless $Tr (K_{j})=0$, while $Tr (K_{i} K_{j})= 2\delta_{ij}$, and the Casimir invariant is $K_{x}^{2} + K_{y}^{2} + K_{z}^{2} = {\bf K}\cdot {\bf K} = 2 I$, where $I$ is the 3x3 identity matrix.
The spin-1 algebra of the angular momenta $K_{x}, K_{y}, K_{z}$ can be further worked out by the standard techniques of constructing the appropriate ladder operators. We denote the common set of eigenalues of ${\mathbf K}^2$ and each of the 
momenta $K_{x}, K_{y}, K_{z}$ by $|1,m\rangle_{x}$, $|1,m\rangle_{y}$, $|1,m\rangle_{y}$, with $m=-1,0,1$. Specifically, we find
\begin{widetext}
	\begin{equation}
	\begin{aligned}
	|1,0\rangle_{x} = \left[\begin{array}{c} 0\\ 0 \\-i \end{array} \right], ~~
	|1,0\rangle_{y} = \left[\begin{array}{c} 1\\ 0\\0 \end{array} \right], ~~
	|1,0\rangle_{z} = \left[\begin{array}{ccc} 0\\ 1 \\0 \end{array} \right], \\
	|1,\pm 1\rangle_{x} = \frac{1}{\sqrt{2}}\left[\begin{array}{c} 1\\ \pm 1 \\ 0\end{array} \right], ~~
	|1,\pm 1\rangle_{y} = \frac{1}{\sqrt{2}}\left[\begin{array}{c} 0\\ \mp i\\ -i\end{array} \right], ~~
	|1,\pm 1 \rangle_{z} = \frac{1}{\sqrt{2}}\left[\begin{array}{ccc} 1\\ 0 \\ \pm i \end{array} \right],	
	\end{aligned}
	\end{equation}
\end{widetext}

Note that the more usual representation $J_{x}, J_{y}, J_{z}$ for spin-1 angular momentum operators,
\begin{equation}
\begin{aligned}
J_{x} = \frac{1}{\sqrt{2}}\left[\begin{array}{ccc} 0&1&0\\ 1&0&1 \\0&1&0 \end{array} \right], ~~~ 
J_{y} = \frac{1}{\sqrt{2}i}\left[\begin{array}{ccc} 0&1&0\\ -1&0&1 \\0&-1&0 \end{array} \right], \nonumber 
\end{aligned}
\end{equation}
and 
\begin{equation}
\begin{aligned}
J_{z} = \left[\begin{array}{ccc} 1&0&0\\ 0&0&0 \\0&0&-1 \end{array} \right],
\end{aligned}
\end{equation}
can be recovered via a transformation $J_{i} = S^{\dag} K_{i} S$, where the unitary $S$ is defined as \cite{Carroll:88}
\begin{equation}
\begin{aligned}
S= \frac{1}{\sqrt{2}} \left[\begin{array}{ccc} 1&0&1\\ 0&\sqrt{2}&0 \\i&0&-i \end{array} \right].
\end{aligned}
\end{equation}

Thus, the saSTIRAP Hamiltonian reads
\begin{equation}
H= \frac{\hbar}{2}\left[\Omega_{01}(t)K_{x} + \Omega_{12}(t)K_{y} + \Omega_{02}(t)K_{z} \right],
\end{equation}
which represents the Hamiltonian of a spin-1 particle with magnetic field components $\left(\Omega_{01}(t),\Omega_{12}(t),\Omega_{02}(t)\right)$.

\section*{SI2: Inhomogenous-broadening and ac Stark shift effects}

We now focus on the energy-shifting terms, which are represented by the inhomogenous-broadening Hamiltonian Eq. (\ref{eq:inh}) in the spin system and by the ac Stark shifts (\ref{eq:acStark}) in the simulator.  In general, any fields detuned from a transition produce not only the Rabi coupling but also a energy level shift. In the spin system, these are produced by the modulation of the coupling and will shift the Larmor frequency of each spin, resulting in inhomogenous broadening. By extensive numerical simulations, we have verified that the effect of these terms is to change the optimal gauge-invariant phase $\Phi$ from the ideal value of $-\pi/2$, as well as to reduce the overall transfer fidelity. While the latter effect is expected, the change in the optimal gauge-invariant can be understood as a result of the slow accumulation of phase differences between the counterdiabatic pulse and the STIRAP during the run of the protocol.

To see why this is the case, consider first a two-photon pulse in the absence of STIRAP. The effective Hamiltonian in the $0 - 2$ subspace is
\begin{equation}
H_{\rm 2 ph} = \frac{\hbar}{2}\begin{bmatrix} \epsilon & \Omega_{02} e^{2 i \tilde{\varphi} + i\pi} \\ \Omega_{02} e^{-2 i \tilde{\varphi} - i \pi} & -\epsilon \end{bmatrix},
\end{equation}
where $\epsilon = (\tilde{\Omega}_{12}^2 -\tilde{\Omega}_{01}^2)/(4 \Delta )$ and $\Omega_{02} = \tilde{\Omega}_{01}\tilde{\Omega}_{12}/(2\Delta )$, where $\Delta = (\omega_{01} - \omega_{12})/2$ is the detuning. The maximum population transfer by a two-photon $\pi$ Rabi pulse is therefore
$\Omega_{02} /\sqrt{\epsilon^2 + \Omega_{02}^2}$, and now using the fact that for the transmon $\tilde{\Omega}_{12} = \sqrt{2} \tilde{\Omega}_{01}$, we obtain for the maximum population 0.943, only 5.7\% below the maximum of 1. Thus, the populations are only mildly affected, but if we estimate the phases $\pm (\epsilon/2 ) t_{\pi}$ acquired by each state $|0\rangle$ and $|2\rangle$ accumulated during the $\pi$ pulse of duration $t_{\pi} = \pi /\sqrt{\Omega_{02}^2 + \epsilon^2}$ we find $\pm \pi/6$, which is not negligible.

Now, when STIRAP is added, the Stokes and pump fields will no longer couple resonantly into the levels. Let us consider the situation in full generality, with both STIRAP and the two-photon pulse applied and with corresponding detunings $\tilde{\delta}_{01} = \tilde{\omega} - \omega_{01}$ , $\delta_{01}^{(\Omega )} = \omega_{01}^{(\Omega )} - \omega_{01}$,  $\tilde{\delta}_{12} = \tilde{\omega} - \omega_{12}$ , $\delta_{12}^{(\Omega )} = \omega_{12}^{(\Omega )} - \omega_{12}$.



Then the full Hamiltonian in the rotating wave approximation for the STIRAP part reads, in matrix form,
\begin{equation}
H = H_{\rm 2ph} + H_{\rm RWA}^{\rm (STIRAP)},
\end{equation}
where
\begin{equation}
H_{\rm 2ph}(t) = \frac{\hbar}{2} \begin{bmatrix} 0 & \tilde{\Omega}_{01}(t) e^{i (\tilde{\delta}_{01} t + \tilde{\varphi})}  & 0 \\
\tilde{\Omega}_{01}(t) e^{-i (\tilde{\delta}_{01} t + \tilde{\varphi})}  & 0 & \tilde{\Omega}_{12}(t) e^{i (\tilde{\delta}_{12} t + \tilde{\varphi})} \\ 0 & \tilde{\Omega}_{12}(t) e^{-i (\tilde{\delta}_{12} t + \tilde{\varphi})}  & 0 \end{bmatrix},
\end{equation}
and
\begin{widetext}
	\begin{equation}
	H_{\rm RWA}^{\rm (STIRAP)}(t) = \frac{\hbar}{2} \begin{bmatrix} 0 &  \Omega_{01}(t) e^{i \left(\delta^{(\Omega)}_{01} t + \phi_{01}\right)} & 0 \\
	\Omega_{01}(t) e^{-i \left(\delta^{(\Omega)}_{01} t + \phi_{01}\right)} & 0 &  \Omega_{12}(t) e^{i \left(\delta^{(\Omega)}_{12} t + \phi_{12}\right)} \\ 0 &  \Omega_{12}(t) e^{-i \left(\delta^{(\Omega)}_{12} t + \phi_{12}\right)} & 0 \end{bmatrix}.
	\end{equation}
\end{widetext}
Here $\tilde{\delta}_{01} = \tilde{\omega} - \omega_{01}$ , $\delta_{01}^{(\Omega )} = \omega_{01}^{(\Omega )} - \omega_{01}$,  $\tilde{\delta}_{12} = \tilde{\omega} - \omega_{12}$ , $\delta_{12}^{(\Omega )} = \omega_{12}^{(\Omega )} - \omega_{12}$.
Next, we notice that the detunings of two-photon pulse are much larger than those of the STIRAP pulse. This allows us to regard the Hamiltonian $H_{\rm 2ph}$ as a fast-variable Hamiltonian, in contrast with $H_{\rm RWA}^{\rm (STIRAP)}$, which can be considered as a slow-variable Hamiltonian.  Next, we apply the formalism of effective Hamiltonians \cite{James2007} by averaging over the fast variables, resulting in
\begin{equation}
H_{\rm av} =  H_{\rm RWA}^{\rm (STIRAP)} + \frac{1}{2} \overline{[H_{\rm 2ph}, V_{\rm 2ph}]},
\end{equation}
where $[ , ]$ denotes the commutator, the overbar denotes the mean, and $V_{\rm 2ph}$ is 
\begin{equation}
V_{\rm 2ph}(t) = \frac{1}{2} \begin{bmatrix} 0 & -\frac{\tilde{\Omega}_{01}(t)}{\tilde{\delta}_{01}} e^{i (\tilde{\delta}_{01} t + \tilde{\varphi})}  & 0 \\
\frac{\tilde{\Omega}_{01}(t)}{\tilde{\delta}_{01}} e^{-i (\tilde{\delta}_{01} t + \tilde{\varphi})}  & 0 & -\frac{\tilde{\Omega}_{12}(t)}{\tilde{\delta}_{12}} e^{i (\tilde{\delta}_{12} t + \tilde{\varphi})} \\ 0 & \frac{\tilde{\Omega}_{12}(t)}{\tilde{\delta}_{12}} e^{-i (\tilde{\delta}_{12} t + \tilde{\varphi})}  & 0 \end{bmatrix}.
\end{equation}

After some algebra we obtain
\begin{widetext}
	\begin{equation}
	H_{\rm av}(t) = \frac{\hbar}{2} \begin{bmatrix} 2 \epsilon_{0}(t) &  \Omega_{01}(t) e^{i \left(\delta^{(\Omega)}_{01} t + \phi_{01}\right)} & \Omega_{02}(t) e^{i \left( 2 \tilde{\varphi} + \pi - \tilde{\delta}_{02}\right)} \\
	\Omega_{01}(t) e^{-i \left(\delta^{(\Omega)}_{01} t + \phi_{01}\right)} & 2 \epsilon_{1}(t) &  \Omega_{12}(t) e^{i \left(\delta^{(\Omega)}_{12} t + \phi_{12}\right)} \\ \Omega_{02}(t) e^{-i \left(2 \tilde{\varphi} + \pi - \tilde{\delta}_{02}\right)} &  \Omega_{12}(t) e^{-i \left(\delta^{(\Omega)}_{12} t + \phi_{12}\right)} & 2 \epsilon_{2}(t) \end{bmatrix} .
	\end{equation}
\end{widetext}
This allows us to identify the energy-level shifts
\begin{equation}
\epsilon_{0}(t) = - \frac{1}{2} \frac{\tilde{\Omega}_{01}(t)^2}{2 \Delta - \tilde{\delta}_{02}} ,
\end{equation}
\begin{equation}
\epsilon_{1}(t) =  \frac{1}{2} \left(\frac{\tilde{\Omega}_{01}(t)^2}{2 \Delta - \tilde{\delta}_{02}} +
\frac{\tilde{\Omega}_{12}(t)^2}{2 \Delta + \tilde{\delta}_{02}}\right) ,
\end{equation}
\begin{equation}
\epsilon_{2}(t) = - \frac{1}{2} \frac{\tilde{\Omega}_{12}(t)^2}{2 \Delta + \tilde{\delta}_{02}} .
\end{equation}
where  $2 \Delta = \omega_{12}-\omega_{01}$ equals the qubit anharmonicity and $\tilde{\omega}$, $\tilde{\delta}_{02} = \omega_{01} + \omega_{12} - 2 \tilde{\omega}$ is the two-photon detuning.
Next, we can eliminate the diagonal terms by the transfomation
\begin{equation}
U_{\chi} (t) = e^{i \int_{0}^{t} \epsilon_{0}(\tau ) d\tau }|0\rangle \langle 0| + e^{i \int_{0}^{t} \epsilon_{1}(\tau ) d\tau }|1\rangle \langle 1|
+ e^{i \int_{0}^{t} \epsilon_{2}(\tau ) d\tau }|2\rangle \langle 2| ,
\end{equation}
This brings the effective Hamiltonian in the form
\begin{widetext}
	\begin{equation}
	H^{(\rm eff)} (t) = \frac{\hbar}{2}\left[\begin{array}{ccc} 0 & \Omega_{01}(t) e^{i \phi_{01}^{\rm (eff)}(t)} & \Omega_{02}(t)e^{i \phi_{02}^{\rm (eff)}(t)} \\ \Omega_{01}(t) e^{-i \phi_{01}^{\rm (eff)}(t)} & 0 & \Omega_{12}(t) e^{-i \phi_{12}^{\rm (eff)}(t)} \\ \Omega_{02}(t) e^{-i \phi_{02}^{\rm (eff)}(t)} & \Omega_{12}(t) e^{-i \phi_{12}^{\rm (eff)}(t)} & 0 \end{array} \right], \label{eq:effective:standard}
	\end{equation}
\end{widetext}
where
\begin{eqnarray}
\phi_{01}^{\rm (eff)}(t) &=& \phi_{01} + \delta^{(\Omega)}_{01} t + \int_{0}^{t} d\tau [\epsilon_{0}(\tau ) - \epsilon_{1}(\tau )] ,\\
\phi_{12}^{\rm (eff)}(t) &=& \phi_{12} + \delta^{(\Omega)}_{12} t + \int_{0}^{t} d\tau [\epsilon_{1}(\tau ) - \epsilon_{2}(\tau )] ,\\
\phi_{02}^{\rm (eff)}(t) &=& 2\tilde{\varphi} + i \pi + \tilde{\delta}_{02} t + \int_{0}^{t} d\tau [\epsilon_{0}(\tau ) - \epsilon_{2}(\tau )] .
\end{eqnarray}
We see that in general these phases are time-dependent; however, the dependence due to the ac Stark shift is a slow dependence that accumulates during the interval from $0$ to $t$.  The critical period of time is when the pulses have high amplitude, since that is when the transfer occurs. In this region, by changing the detunings
$\delta^{(\Omega)}_{01}$, $\delta^{(\Omega)}_{12}$, and $\tilde{\delta}_{02}$ we can find a situation in which during the transfer is optimal.

\begin{acknowledgments} 
	We are grateful to Sergey Danilin for sample preparation and assistance with the cryogenics and the measurements and to Shruti Dogra and Pieter W. Claeys for useful comments on the manuscript.
	This work used the cryogenic facilities of the Low Temperature Laboratory at Aalto University. We acknowledge 
	funding from the European Union’s Horizon 2020 research and innovation programs  under grant agreement No. 862644 (FET-Open QUARTET).
	We are grateful for financial support from the Foundational Questions Institute Fund (FQXi) under Grant No. FQXi-IAF19-06, from the V\"aisal\"a Foundation,
	and from the Academy of Finland through 
	projects 263457, 250280, and 328193. This work was performed under the “Finnish Center of Excellence in Quantum Technology” of the Academy of Finland (project 312296).
\end{acknowledgments}

\bibliography{ref_saSTIRAP}




\end{document}